\documentclass[aps,prc,twocolumn,superscriptaddress,floatfix,amsmath,amssymb,nofootinbib,longbibliography]{revtex4-1}
\usepackage[utf8]{inputenc}
\usepackage[T1]{fontenc}

\usepackage{xcolor}
\usepackage{graphicx}

\usepackage{scalefnt}
\usepackage{times,txfonts}

\usepackage{amsmath,nicefrac}
\usepackage{amssymb}
\usepackage{amsfonts}
\usepackage{mathtools}
\usepackage{tabularx}
\usepackage{booktabs}

\usepackage{braket}

\usepackage[citecolor=blue,linkcolor=blue,urlcolor=blue]{hyperref}
\usepackage[ocgcolorlinks]{ocgx2}

\newcommand{\clebsch}[6]{\mathcal{C}_{#1 #2 #3 #4}^{#5 #6}}

\newcommand{\sixj}[6]{\begingroup\setlength{\arraycolsep}{0.2em}\begin{Bmatrix} #1 & #2 & #3 \\ #4 & #5 & #6 \end{Bmatrix}\endgroup}
\newcommand{\ninej}[9]{\begingroup\setlength{\arraycolsep}{0.2em}\begin{Bmatrix} #1 & #2 & #3 \\ #4 & #5 & #6 \\ #7 & #8 & #9 \end{Bmatrix}\endgroup}

\newcommand{\qncm}[1]{{\ensuremath{{#1}_\text{CM}}}}
\newcommand{\qnrel}[1]{{\ensuremath{\bar {#1}}}}
\newcommand{\qntot}[1]{{\ensuremath{{#1}_\text{tot}}}}
\newcommand{\qnjac}[1]{\ensuremath{#1}}
\newcommand{\mproj}[1]{\ensuremath{M_{#1}}}

\newcommand{\pr}{\prime}

\newcommand{\la}{\langle}
\newcommand{\ra}{\rangle}

\newcommand{\cm}{{\text{CM}}}

\newcommand{\ai}{{\emph{ab initio}}}

\allowdisplaybreaks

\begin{document}

\title{Normal ordering of three-nucleon interactions for \ai\ calculations of heavy nuclei}

\author{K.~Hebeler}
\email{kai.hebeler@physik.tu-darmstadt.de}
\affiliation{Technische Universit\"at Darmstadt, Department of Physics, 64289 Darmstadt, Germany}
\affiliation{ExtreMe Matter Institute EMMI, GSI Helmholtzzentrum f\"ur Schwerionenforschung GmbH, 64291 Darmstadt, Germany}
\affiliation{Max-Planck-Institut f\"ur Kernphysik, Saupfercheckweg 1, 69117 Heidelberg, Germany}

\author{V.~Durant}
\email{vdurant@uni-mainz.de}
\affiliation{Institut für Kernphysik and PRISMA+ Cluster of Excellence,
Johannes Gutenberg-Universität Mainz, D-55128 Mainz, Germany}

\author{J.~Hoppe}
\email{jhoppe@theorie.ikp.physik.tu-darmstadt.de}
\affiliation{Technische Universit\"at Darmstadt, Department of Physics, 64289 Darmstadt, Germany}
\affiliation{ExtreMe Matter Institute EMMI, GSI Helmholtzzentrum f\"ur Schwerionenforschung GmbH, 64291 Darmstadt, Germany}

\author{M.~Heinz}
\email{mheinz@theorie.ikp.physik.tu-darmstadt.de}
\affiliation{Technische Universit\"at Darmstadt, Department of Physics, 64289 Darmstadt, Germany}
\affiliation{ExtreMe Matter Institute EMMI, GSI Helmholtzzentrum f\"ur Schwerionenforschung GmbH, 64291 Darmstadt, Germany}
\affiliation{Max-Planck-Institut f\"ur Kernphysik, Saupfercheckweg 1, 69117 Heidelberg, Germany}

\author{A.~Schwenk}
\email{schwenk@physik.tu-darmstadt.de}
\affiliation{Technische Universit\"at Darmstadt, Department of Physics, 64289 Darmstadt, Germany}
\affiliation{ExtreMe Matter Institute EMMI, GSI Helmholtzzentrum f\"ur Schwerionenforschung GmbH, 64291 Darmstadt, Germany}
\affiliation{Max-Planck-Institut f\"ur Kernphysik, Saupfercheckweg 1, 69117 Heidelberg, Germany}

\author{J.~Simonis}
\email{simonis@theorie.ikp.physik.tu-darmstadt.de}
\affiliation{Institut für Kernphysik and PRISMA+ Cluster of Excellence,
Johannes Gutenberg-Universität Mainz, D-55128 Mainz, Germany}

\author{A.~Tichai}
\email{alexander.tichai@physik.tu-darmstadt.de}
\affiliation{Technische Universit\"at Darmstadt, Department of Physics, 64289 Darmstadt, Germany}
\affiliation{ExtreMe Matter Institute EMMI, GSI Helmholtzzentrum f\"ur Schwerionenforschung GmbH, 64291 Darmstadt, Germany}
\affiliation{Max-Planck-Institut f\"ur Kernphysik, Saupfercheckweg 1, 69117 Heidelberg, Germany}

\begin{abstract}
Three-nucleon (3N) interactions are key for an accurate solution of the nuclear many-body problem. However, fully taking into account 3N forces constitutes a computational challenge and hence approximate treatments are commonly employed. The method of normal ordering has proven to be a powerful tool that allows to systematically include 3N interactions in an efficient way, but traditional normal-ordering frameworks require the representation of 3N interactions in a large single-particle basis, typically necessitating a truncation of 3N matrix elements. While this truncation has only a minor impact for light and medium-mass nuclei, its effects become sizable for heavier systems and hence limit the scope of \textit{ab initio} calculations. In this work, we present a novel normal-ordering framework that allows to circumvent this limitation by performing the normal ordering directly in a Jacobi basis. We discuss in detail the new framework, benchmark it against established results, and present calculations for ground-state energies and charge radii of heavy nuclei, such as $^{132}$Sn and $^{208}$Pb.
\end{abstract}

\maketitle

\section{Introduction}

The inclusion of three-body forces in nuclear Hamiltonians is crucial to obtain a realistic description of the structure of finite nuclei and properties of dense matter~\cite{Hamm13RMP,Hebe15ARNPS,Hebe203NF}.
However, the full inclusion of three-body operators is computationally very challenging due to a steep increase in the number of matrix elements when using large model spaces, as is required for converged calculations of heavier nuclei. Normal ordering (NO) is a powerful and well-established method that allows to transform a given Hamiltonian in an exact way, such that contributions from 3N interactions can be incorporated to good approximation at the computational cost of two-body interactions. This so-called normal-ordered two-body approximation (NO2B) has become the standard tool in state-of-the-art calculations of finite nuclei~\cite{Hage07CC3N,Otsu10Ox,Roth12NCSMCC3N,Bind14CCheavy,Gebr16MR,Simo17SatFinNuc,Stro21atomicNucl} and nuclear matter~\cite{Hebe10nmatt,Holt10ddnn,Carb13SCGF3B,Dris16asym}.

However, established NO frameworks for \textit{ab initio} calculations of nuclei require the representation of 3N interactions in a given single-particle basis as an intermediate step.
Due to the rapid increase of the three-body basis dimension, this step necessitates the introduction of additional truncations for a given single-particle model space. Typically, a cut on the allowed three-body energy quantum numbers is applied. For studies of light and medium-mass nuclei the impact of the additional truncation is small, and for low resolution interactions calculations converge rapidly at moderate truncation values. For heavier systems this cut becomes significant, and the uncertainty due to these effects has constrained \textit{ab initio} calculations to masses $A \lesssim 100$.
Only recently, extensions to larger three-body spaces have become available for the NO2B approximation, allowing converged calculations for heavy nuclei based on soft nuclear interactions~\cite{Miya21E3max,Hu21Pb208skin}.

In this work, we present a novel NO framework that is formulated directly in a partial-wave-decomposed Jacobi-momentum basis~\cite{Gloe83qmfewbod,Hebe15N3LOpw,Hebe203NF}, in which 3N interaction matrix elements are most commonly stored. Performing the NO in this basis avoids the need to represent the 3N interaction in a single-particle basis at any point of the calculation and hence fully circumvents memory limitations associated with the single-particle formulation of the NO2B approximation~\cite{Roth14SRG3N}. The new Jacobi NO framework yields an effective two-body interaction that explicitly depends on the center-of-mass (CM) momentum, characterized by an extended set of quantum numbers.
In this work, we present the details of the new framework and carefully benchmark it against the existing NO implementation for a selected set of closed-shell nuclei with a large range of mass numbers from $^{16}$O up to $^{208}$Pb.

This work is structured as follows:
In Sec.~\ref{sec:spno}, we revisit the concept of NO in its single-particle formulation.
Section~\ref{sec:jacno} introduces the novel Jacobi NO framework.
Results and numerical benchmarks are presented in Sec.~\ref{sec:res}.
Finally, we summarize and conclude with future perspectives in Sec.~\ref{sec:outlook}.

\section{Normal ordering:\\Traditional formulation}
\label{sec:spno}

\subsection{Reference states}

Basis-expansion methods for nuclei employ an $A$-body reference state $\ket{\Phi}$ as a starting point for the correlation expansion~\cite{Dick04PPNP,Herg16PR,Hage14RPP,Herg20review,Tichai2020review}.
The reference state provides the many-body method with a qualitatively correct starting point for an expansion of the state of interest.
In this work, we use a spherical $A$-particle Slater-determinant reference state
\begin{equation}
|\Phi \rangle = \prod_{i=1}^A a^\dagger_i \ket{0} \, ,
\end{equation}
where $|0 \ra$ denotes the physical vacuum.
The $A$ particles in the system occupy orthonormal single-particle states
\begin{equation}
a^\dagger_p \left| 0 \right>= \ket{\varphi_p} = \ket{\varphi_{n_p l_p j_p m_{j_p} m_{t_p}}} \, .
\label{eq:HO_single_particle_coupled}
\end{equation}
Here $n_p$ denotes the radial quantum number, and $l_p$ denotes the single-particle orbital angular-momentum quantum number coupled with the spin $s_p = \tfrac{1}{2}$ to the total angular-momentum quantum number $j_p$ and its projection $m_{j_p}$. The isospin is $t_p = \tfrac{1}{2}$ and the projection quantum number $m_{t_p}$ denotes proton $(m_{t_p} = +\tfrac{1}{2})$ and neutron states $(m_{t_p} = -\tfrac {1}{2})$. Since our reference state is a single Slater determinant, states occupied in the reference state (i.e., $\ket{\varphi_i}$ with $i\leq A$) have an associated occupation number $\tilde{n}_p =1$.\footnote{The occupation number $\tilde{n}_i$ should not be confused with the radial quantum number $n_i$.} The occupation number of the remaining states in the computational basis is given by $\tilde{n}_i = 0$.

Starting from an initial set of spherical harmonic oscillator (HO) orbitals, one can construct a new set of orbitals, e.g., by solving the Hartree-Fock (HF) equations as is typically done in this work. As long as the HF solution is obtained in a symmetry-restricted way, the single-particle transformation will only mix radial quantum numbers, allowing the HF orbitals to be re-expressed in terms of HO orbitals as
\begin{equation}
\label{eq:HF_ref_state}
\ket{\varphi_{n_i l_i j_i m_{j_i} m_{t_i}}} = \sum_{n'_i} C_{l_i j_i m_{j_i} m_{t_i}} (n_i,n'_i) \ket{\varphi_{n'_i l_i j_i m_{j_i} m_{t_i}}}_{\text{HO}} \, .
\end{equation}
In the future, it will also be interesting to study alternative single-particle bases, such as natural orbitals, based on the eigenstates of a perturbatively improved density matrix (see, e.g., Refs.~\cite{Tich19NatNCSM,Hopp20NAT,Fasa20NATCI}).

\subsection{Single-particle formulation}

Once an appropriate reference state has been determined,
normal-ordered matrix elements for all operators of interest constitute the fundamental input of basis-expansion many-body frameworks~\cite{Dick04PPNP,Hage14RPP,Herg16PR,Herg20review,Tichai2020review}.
The normal-ordered Hamiltonian is commonly written as
\begin{align}
H &= 
E_0 
+\sum_{pq} f_{pq} : a_{p}^{\dagger} a_{q}:
+\frac{1}{4} \sum_{pqrs} \Gamma_{pqrs} : a_{p}^{\dagger} a_{q}^{\dagger} a_{s} a_{r}: \nonumber \\
&\quad+\frac{1}{36} \sum_{pqrstu} W_{pqrstu} : a_{p}^{\dagger} a_{q}^{\dagger} a_{r}^{\dagger} a_{u} a_{t} a_{s}: \, ,
\end{align} 
where $E_0 = \braket{\Phi | H | \Phi}$ is the reference-state expectation value of the Hamiltonian, $f_{pq}$, $\Gamma_{pqrs}$, and $W_{pqrstu}$ are the normal-ordered
one\nobreakdash-, two\nobreakdash-, and three-body matrix elements of the Hamiltonian, and the colons $: \ldots :$ indicate normal-ordered products of fermion creation and annihilation operators.

These matrix elements must be computed from the ``free-space'' Hamiltonian
\begin{align}
H = T + V_{\text{NN}} + V_{\text{3N}} \, ,
\end{align}
which is typically represented in a single-particle basis as an intermediate step:
\begin{subequations}
\begin{align}
T &= \sum_{pq} \left< p | T | q \right> a^{\dagger}_p a_q \, ,  \\
V_{\text{NN}} &= \frac{1}{4} \sum_{pqrs} \bigl< pq | V_{\text{NN}}^{\text{as}} | rs \bigr> \: a_p^{\dagger} a_q^{\dagger} a_s a_r \, , \\
V_{\text{3N}} &= \frac{1}{36} \sum_{pqrstu} \bigl< pqr | V_{\text{3N}}^{\text{as}} | stu \bigr> \: a_p^{\dagger} a_q^{\dagger} a_r^{\dagger} a_u a_t a_s \, .
\end{align}
\end{subequations}
Here $\bigl< pq | V_{\text{NN}}^{\text{as}} | rs \bigr>$ and $\bigl< pqr | V_{\text{3N}}^{\text{as}} | stu \bigr>$ denote the
antisymmetrized matrix elements of the NN and 3N interactions, respectively.
The normal-ordered Hamiltonian matrix elements can be expressed in terms of these free-space single-particle matrix elements as
\begin{subequations}
\begin{align}
E_0 &= \sum_p \tilde{n}_p \bigl< p | T | p \bigr> + \frac{1}{2} \sum_{pq} \tilde{n}_p \tilde{n}_q \bigl< pq | V_\text{NN}^{\text{as}} | pq \bigr> \notag \\
& \quad + \frac{1}{6} \sum_{pqr} \tilde{n}_p \tilde{n}_q \tilde{n}_r \bigl< pqr | V_\text{3N}^{\text{as}} | pqr \bigr> \, , \label{eq:normord_terms_HF_0b}\\  
f_{pq} &= \bigl< p | T | q \bigr> + \sum_{r} \tilde{n}_r \bigl< p r | V_{\text{NN}}^{\text{as}} | qr \bigr> \notag \\
& \quad + \frac{1}{2} \sum_{rs} \tilde{n}_r \tilde{n}_s \bigl< prs | V_{\text{3N}}^{\text{as}} | qrs \bigr> \, , \label{eq:normord_terms_HF_1b}\\
\Gamma_{pqrs}  &= \bigl< pq | V_{\text{NN}}^{\text{as}} | rs \bigr> + \sum_t \tilde{n}_t \bigl< pqt  | V_{\text{3N}}^{\text{as}} | rst \bigr> \, , \label{eq:normord_terms_HF_2b}\\
W_{pqrstu} &= \bigl< pqr | V_{\text{3N}}^{\text{as}} | stu \bigr>\, ,
\label{eq:normord_terms_HF}
\end{align} 
\end{subequations}
with the reference-state occupation numbers $\tilde{n}_p$.

In this single-particle NO scheme,
the computational and storage cost is dominated by the three-body matrix elements $\bigl< pqr | V_{\text{3N}}^{\text{as}} | stu \bigr>$.
For typical single-particle basis truncations, characterized by $e_{\text{max}} = (2 n + l)_{\text{max}} \sim \text{14--16}$,
the storage cost of three-body matrix elements without any additional truncations is tens or hundreds of TB (see Fig.~\ref{fig:dims}).
This necessitates the additional truncation of the basis of three-body states $\left| pqr \right>$,
and one such established truncation discards states based on their three-body energy quantum number
\begin{equation}
E^{(3)} = e_p + e_q + e_r \leq E^{(3)}_{\text{max}}\,.
\end{equation}
The basis dimension for the representation of three-body operators grows rapidly with increasing $E^{(3)}_{\text{max}}$ values. 
In the NO2B approximation the normal-ordered Hamiltonian is truncated at the two-body level, i.e.,
\begin{equation}
H = E_0 +\sum_{pq} f_{pq} : a_{p}^{\dagger} a_{q}:+\frac{1}{4} \sum_{pqrs} \Gamma_{pqrs} : a_{p}^{\dagger} a_{q}^{\dagger} a_{s} a_{r}:\,,
\end{equation}
and the residual three-body contributions from $W_{ijklmn}$ are neglected. The NO2B approximation has been used successfully in various many-body methods.
In fact, many-body calculations including explicit three-body operators are presently limited to light nuclei using methods that do not rely on normal ordering~\cite{Barr13PPNP,Carl15RMP,Lynn19QMC}.
For basis-expansion methods for nuclei, the inclusion of residual three-body interactions has been constrained to small model spaces~\cite{Hage07CC3N,Hein21IMSRG3}
or to studies with softened 3N interactions that have shown their effects to be small~\cite{Roth12NCSMCC3N,Bind13expl3NCCSD,Bind13expl3NLCCSDT,Bind14CCheavy}.

In practice the evaluation of the 3N contribution to the normal-ordered two-body Hamiltonian $\Gamma$ in Eq.~\eqref{eq:normord_terms_HF_2b},
\begin{equation}
\Gamma^{\text{3N}}_{pqrs} \equiv \sum_t \tilde{n}_t \bigl< pqt | V_{\text{3N}}^{\text{as}} | rst \bigr>\,,
\label{eq:NO_contraction}
\end{equation}
constitutes the computationally most challenging step. It formally requires a representation of the 3N interaction in a single-particle basis that is sufficiently large for a well-converged many-body calculation. Lower-rank contributions that enter the normal-ordered one-body and zero-body parts in Eqs.~\eqref{eq:normord_terms_HF_0b} and \eqref{eq:normord_terms_HF_1b}, respectively, can be easily computed from $\Gamma^{\text{3N}}_{pqrs}$.
For many years $E^{(3)} \le E^{(3)}_{\text
{max}} \simeq 16$ was the state of the art for \ai{} many-body calculations.
Only recently, the size of the three-body basis could be significantly increased to $E^{(3)}_\text{max}=28$ at the NO2B level in Ref.~\cite{Miya21E3max} by storing only a subset of matrix elements needed for NO. This made it possible to obtain converged results for soft interactions in the region of $^{132}$Sn~\cite{Miya21E3max} and enabled first \ai{} studies of $^{208}$Pb~\cite{Hu21Pb208skin}.

\section{Normal ordering:\\Jacobi basis formulation}
\label{sec:jacno}

In this section, we discuss how NO can be formulated in the plane-wave Jacobi basis. This circumvents the storage of three-body matrix elements in a single-particle basis at any point and thus completely avoids the $E^{(3)}_{\text{max}}$ cut. Instead, the Jacobi NO framework introduces truncations in the CM and relative partial-wave angular momenta that effectively define the basis dimension. These truncations, however, turn out to be more robust than the $E^{(3)}_{\text{max}}$ cut for heavy systems and allow for significant storage savings compared to the traditional NO formulation. The effective two-body interaction is subsequently transformed to the single-particle basis, yielding a suitable input for many-body methods.

\subsection{Jacobi basis formulation}

For the discussion of the new framework we start by representing the two-body matrix elements of $\Gamma^{\text{3N}}$ in a momentum-space single-particle basis of the form
\begin{equation}
\left| \tilde{\mathbf{k}}_1 \tilde{\mathbf{k}}_2 \right> \quad \text{with} \quad \left| \tilde{\mathbf{k}}_i \right> \equiv \left| \mathbf{k}_i m_{s_i} m_{t_i} \right> \, .
\end{equation}
In this basis Eq.~\eqref{eq:NO_contraction} can be written more explicitly in the form
\begin{align}
\left< \tilde{\mathbf{k}}'_1 \tilde{\mathbf{k}}'_2 \left| \Gamma^{\text{3N}} \right| \tilde{\mathbf{k}}_1 \tilde{\mathbf{k}}_2 \right> &= 
\int \frac{d \mathbf{k}_3 d \mathbf{k}'_3}{(2 \pi)^6} \sum_{m_{s_3} m'_{s_3}}  \sum_{m_{t_3} m'_{t_3}} \nonumber \\ & \times \rho (\tilde{\mathbf{k}}_3, \tilde{\mathbf{k}}'_3) \left< \tilde{\mathbf{k}}'_1 \tilde{\mathbf{k}}'_2 \tilde{\mathbf{k}}'_3 \left| V^{\text{as}}_{\text{3N}} \right| \tilde{\mathbf{k}}_1 \tilde{\mathbf{k}}_2 \tilde{\mathbf{k}}_3 \right> \, ,
\label{eq:Gamma_momentum_sp}
\end{align}
with the density matrix
\begin{equation}
\rho (\tilde{\mathbf{k}}_3, \tilde{\mathbf{k}}'_3) = 
\delta_{m_{t_3} m'_{t_3}} \sum_{n_3 l_3 j_3 m_{j_3}} \tilde{n}_3  \left< \tilde{\mathbf{k}}'_3 \left| \varphi_{n_3 l_3 j_3 m_{j_3} m_{t_3}} \right> \left< \varphi_{n_3 l_3 j_3 m_{j_3} m_{t_3}} \right| \tilde{\mathbf{k}}_3 \right> .
\end{equation}
Note that the density matrix is diagonal in the isospin projection quantum number $m_{t_3}$ but not in the spin projection quantum number $m_{s_3}$.
The reference-state orbitals in the single-particle momentum-space basis take the explicit form
\begin{equation}
\left< \mathbf{k}_i m_{s_i} m_{t_i} \left| \varphi_{n_i l_i j_i m_{j_i} m_{t_i}} \right. \right> = \sum_{m_{l_i}} \mathcal{C}_{l_i m_{l_i} \tfrac{1}{2} m_{s_i}}^{j_i m_{j_i}} Y_{l_i m_{l_i}} (\hat{\mathbf{k}}_i) \: \varphi_{n_i l_i j_i m_{j_i} m_{t_i}} (k_i) \, ,
\end{equation}
with the Clebsch-Gordan coefficients $\mathcal{C}_{l m_{l} 1/2 \ m_{s}}^{j m_{j}}$ coupling the single-particle orbital angular momentum and spin to the total angular momentum, the spherical harmonics $Y_{l m_{l}}$, and the angular orientation $\hat{\mathbf{k}}$ and modulus $k = |\mathbf{k}|$ of the vector $\mathbf{k}$.

For practical calculations Eq.~\eqref{eq:Gamma_momentum_sp} is not very useful
due to the complexity and redundancy of 3N interactions
when using the single-particle momentum representation.
Significant benefits can be obtained by exploiting the symmetries of 3N forces, e.g., Galilean, rotational, and isospin invariance. In particular, we explicitly make
use of Galilean invariance of the nuclear interactions by representing it in terms of the relative
and CM momenta, with the interaction being independent of the three-body CM momentum $\mathbf{P}_\text{3N}$. We therefore switch to a Jacobi momentum-space representation by defining the following Jacobi momenta~\cite{Gloe83qmfewbod}:
\begin{align}
\mathbf{p} = \frac{1}{2} \left( \mathbf{k}_1 - \mathbf{k}_2 \right) \, , \quad  \mathbf{q} = \frac{2}{3} \left[ \mathbf{k}_3 - \frac{1}{2} \left( \mathbf{k}_1 + \mathbf{k}_2 \right) \right] \, ,
\end{align}
and the corresponding two- and three-body CM momenta
\begin{subequations}
\begin{align}
\mathbf{P} &= \mathbf{k}_1 + \mathbf{k}_2 \, , \\
\mathbf{P}_{\text{3N}} &= \mathbf{k}_1 + \mathbf{k}_2 + \mathbf{k}_3 =  \mathbf{k}'_1 + \mathbf{k}'_2 + \mathbf{k}'_3 \, ,
\end{align}
\label{eq:cmmomenta}
\end{subequations}
respectively. Including spin and isospin quantum numbers the states are given by 
\begin{subequations}
\begin{align}
\left| \tilde{\mathbf{p}} \right> &= \left| \mathbf{p} S M_S T M_T\right>\, , \\  
\left| \tilde{\mathbf{q}} \right> &= \left| \mathbf{q} m_{s} m_{t} \right> 
=\left| \mathbf{q} m_{s_3} m_{t_3} \right>\, ,
\end{align}
\end{subequations}
where, $S$ ($T$) denotes the two-body spin (isospin) and its projection $M_S$ ($M_T$), and $m_s$ ($m_t$) the spin (isospin) projection of the third particle. The latter are identical to $m_{s_3}$ and $m_{t_3}$ of $\ket{\tilde{\mathbf{k}}_3}$.
From Eqs.~\eqref{eq:cmmomenta} it follows that $\mathbf{k}_3 = 3/2 \mathbf{q} + \mathbf{P}/2$. In this representation Eq.~(\ref{eq:Gamma_momentum_sp}) can be rewritten by expressing the 3N interaction $V^\text{as}_\text{3N}$ in the Jacobi basis. The resulting interaction is independent of $\mathbf{P}_\text{3N}$ and only depends on four momenta instead of six as in the single-particle representation
\begin{align}
\left< \tilde{\mathbf{p}}' \mathbf{P}' \left| \Gamma^{\text{3N}} \right| \tilde{\mathbf{p}} \mathbf{P} \right> &= 
\int \frac{d \mathbf{k}_3 d \mathbf{k}'_3}{(2 \pi)^6} \sum_{m_{s_3} m'_{s_3}}  \sum_{m_{t_3} m'_{t_3}} \rho (\tilde{\mathbf{k}}_3, \tilde{\mathbf{k}}'_3) \nonumber \\
& \times \left< \tilde{\mathbf{p}}' \tilde{\mathbf{q}}' \left| V^{\text{as}}_{\text{3N}} \right| \tilde{\mathbf{p}} \tilde{\mathbf{q}} \right>  (2 \pi)^3 \delta (\mathbf{P} + \mathbf{k}_3 - \mathbf{P}' - \mathbf{k}'_3) \, .
\label{eq:Gamma_momentum_rel}
\end{align}
In the above equation, we have explicitly factored out the trivial dependence of the 3N
interaction matrix elements on the three-body CM momentum $\mathbf{P}_\text{3N} = \mathbf{P} + \mathbf{k}_3$ and
represented the two-body operator $\Gamma^{\text{3N}}$ as a function of the
two-body Jacobi and CM momenta. Note that the two-body
CM momentum $\mathbf{P}$ is in general not conserved since
$\mathbf{k}_3 \neq \mathbf{k}'_3$, and consequently, the resulting two-body interaction $\Gamma^{\text{3N}}$ will depend on $\mathbf{P}$ and $\mathbf{P}'$.
This is in contrast to NO with
respect to momentum eigenstates, as is the case for nuclear matter~\cite{Hebe10nmatt,Dris17MCshort}. 

\subsection{Partial-wave decomposition}

The feasibility of our approach is based on the use of a partial-wave decomposed $\mathcal{J} \mathcal{T}$-coupled basis
\begin{equation}
\label{eq:alpha_basis}
    \ket{p q [(LS)J(ls)j]\mathcal{J}(T t)\mathcal{T}} \equiv \ket{p q \alpha} \,,
\end{equation}
where $p=|\mathbf{p}|$ and $q=|\mathbf{q}|$ define the modulus of the Jacobi momenta, and $\alpha$ refers to all partial-wave quantum numbers:
orbital angular momentum $L$,
spin $S$,
total angular momentum $J$,
and the isospin $T$ of the two-body subsystem of particles 1 and 2 with Jacobi momentum $p$;
orbital angular momentum $l$,
spin $s=1/2$,
total angular momentum $j$,
and the isospin $t=1/2$ of particle 3 relative to the two-body subsystem;
and the total three-body angular momentum $\mathcal{J}$
and total three-body isospin $\mathcal{T}$.
These two three-body quantum numbers also have associated projections $M_{\mathcal{J}}$ and $M_{\mathcal{T}}$, respectively.
However, due to the rotational and isospin invariance of 3N interactions,
the 3N Jacobi matrix elements are diagonal in
$\mathcal{J}$ and $\mathcal{T}$
and independent of $M_{\mathcal{J}}$ and $M_{\mathcal{T}}$.
Therefore, it is sufficient to guarantee that these conditions between the bra and ket states are fulfilled and drop the trivial dependence on the three-body projection quantum numbers.

As a next step, we perform a partial-wave decomposition of the 3N interaction in the Jacobi momentum-space basis (see, e.g., Refs.~\cite{Gloe83qmfewbod,Hebe15N3LOpw,Hebe203NF} for details) with the goal of obtaining a partial-wave decomposed two-body interaction $\Gamma^\text{3N}$.
Using the notation of Eq.~\eqref{eq:alpha_basis}, we first decompose the plane-wave Jacobi basis states into partial waves
\begin{widetext}
\begin{equation}
    \ket{\tilde{\mathbf{p}}\tilde{\mathbf{q}}} = \sum_{L M_L}\sum_{ l m_l }
    Y^{*}_{L M_L}(\hat{\mathbf{p}})
    Y^{*}_{l m_l}(\hat{\mathbf{q}}) 
    \sum_{ J M_J }\sum_{ j m_j}
    \clebsch{L}{M_L}{S}{M_S}{J}{M_J}
    \clebsch{l}{m_l}{\frac{1}{2}}{m_s}{j}{m_j} 
    \sum_{\mathcal{J} M_{\mathcal{J}} }
    \sum_{ \mathcal{T} M_{\mathcal{T}}}
    \clebsch{J}{M_J}{j}{m_j}{\mathcal{J}}{M_{\mathcal{J}}}
    \clebsch{T}{M_T}{\frac{1}{2}}{m_t}{\mathcal{T}}{M_{\mathcal{T}}}
    \ket{p q [(LS)J(ls)j]\mathcal{J}(T t)\mathcal{T}} \, .
    \label{eq:state_PW}
\end{equation}
\end{widetext}

In the following, we will make use of Eq.~(\ref{eq:state_PW}) to represent Eq.~\eqref{eq:Gamma_momentum_rel} in a partial-wave representation. Since the 3N contribution to the normal-ordered two-body interaction breaks Galilean invariance, we have to explicitly incorporate additional quantum
numbers that describe the two-body CM dependence. In the following, we indicate quantum numbers of the two-body basis by an overline to distinguish them from the three-body basis and use a
collective index $\gamma$ for the two-body states. The modified partial-wave decomposed two-body state, including the additional quantum numbers $\bar{L}_\text{CM}$ and $\bar{J}_\text{tot}$, is defined as
\begin{equation}
\bigl| p P \gamma \bigr> \equiv \left| p P [ (\qnrel{L} \qnrel{S}) \qnrel{J} \qncm{\bar{L}} ] \qntot{\bar{J}} \mproj{\qntot{\bar{J}}} \bar{T} M_{\bar{T}} \right> \,,
\label{eq:gamma_basis}
\end{equation} 
with the total two-body angular momentum $\bar{J}_\text{tot}$ obtained by coupling the angular momentum $\bar{J}$ and the two-body CM orbital angular momentum $\bar{L}_\text{CM}$. 
As long as the reference state conserves rotational invariance, the effective two-body interaction $\Gamma^\text{3N}$ transforms like a scalar under rotations in spin and space, is diagonal in the
total angular momentum $\bar{J}_\text{tot}$, does not depend on its projection  $M_{\bar{J}_\text{tot}}$, and is diagonal in the isospin projection $M_{\bar{T}}$.

By inserting the partial-wave expansion, Eq.~(\ref{eq:state_PW}), in Eq.~(\ref{eq:Gamma_momentum_rel}) and projecting the matrix
elements of $\Gamma^{\text{3N}}$ onto the partial-wave states, taking into account the additional CM dependence in 
Eq.~(\ref{eq:gamma_basis}), we obtain the following expression of the normal-ordered 3N contribution in the relative momentum-space basis,
\begin{widetext}
\begin{align}
\begin{split}
\bigl< p' P' \gamma' \left| \Gamma^{\text{3N}} \right| p P \gamma \bigr> &= 
\frac{1}{2 \bar{J}_{\text{tot}} + 1} \sum_{M_{\bar{J}_{\text{tot}}}} 
\sum_{ \substack{ M_{\bar{L}} M_{\bar{S}} M_{\bar{J}} M_{\bar{L}_{\cm}} \\  M'_{\bar{L}} M'_{\bar{S}} M'_{\bar{J}} M^{\pr}_{\bar{L}_{\cm}}}} 
\clebsch{\qnrel{L}}{\mproj{\qnrel{L}}}{\qnrel{S}}{\mproj{\qnrel{S}}}{\qnrel{J}}{\mproj{\qnrel{J}}}
\clebsch{\qnrel{L}^\pr}{M'_{\qnrel{L}}}{\qnrel{S}^\pr}{M'_{\qnrel{S}}}{\qnrel{J}^\pr}{M'_{\qnrel{J}}}
\clebsch{\qnrel{J}}{\mproj{\qnrel{J}}}{\qncm{\bar{L}}}{M_{\bar{L}_{\cm}}}{\qntot{\bar{J}}}{\mproj{\qntot{\bar{J}}}}
\clebsch{\qnrel{J}^\pr}{M'_{\qnrel{J}}}{\qncm{\bar{L}}^\pr}{M^{\pr}_{\bar{L}_{\cm}}}{\qntot{\bar{J}}}{\mproj{\qntot{\bar{J}}}} \\
& \quad \times \int d\hat{\mathbf{P}} \int d \hat{\mathbf{P}}' Y_{\bar{L}_{\cm} M_{\bar{L}_{\cm}}} (\hat{\mathbf{P}})\, Y^*_{\bar{L}'_{\cm} M'_{\bar{L}_{\cm}}} (\hat{\mathbf{P}}') \int \frac{d \mathbf{k}_3}{( 2\pi)^3} \\
& \quad\times \sum_{\substack{ n_3 l_3 j_3 m_{j_3} m_{t_3}\\ m_{l_3} m'_{l_3} m_{s_3} m'_{s_3} }} \clebsch{l_3}{m_{l_3}}{\tfrac{1}{2}}{m_{s_3}}{j_3}{m_{j_3}} \clebsch{l_3}{m'_{l_3}}{\tfrac{1}{2}}{m'_{s_3}}{j_3}{m_{j_3}} \varphi_{n_3 l_3 j_3 m_{j_3} m_{t_3}} (k_3)\, \varphi_{n_3 l_3 j_3 m_{j_3} m_{t_3}} (k'_3) Y_{l_3 m'_{l_3}} (\hat{\mathbf{k}}'_3) Y_{l_3 m_{l_3}}^* (\hat{\mathbf{k}}_3) \\
& \quad\times \sum_{\mathcal{J} M_{\mathcal{J}}} \sum_{\mathcal{T} M_{\mathcal{T}}} \sum_{L L' l l'}  \sum_{J J' jj^\pr} \sum_{M_LM'_L} \sum_{m_l m'_l} \sum_{M_J M'_J} \sum_{ m_j m'_j}  
\delta_{L \bar{L}} \delta_{M_L M_{\bar{L}}} \delta_{L' \bar{L}'} \delta_{M'_L M'_{\bar{L}}} Y^*_{l m_l} (\hat{\mathbf{q}}) Y_{l' m'_l} (\hat{\mathbf{q}}') \\
& \quad\times 
\clebsch{L}{M_L}{\qnrel{S}}{\mproj{\qnrel{S}}}{\qnjac{J}}{\mproj{\qnjac{J}}}
\clebsch{L'}{M'_L}{\qnrel{S}^\pr}{\mproj{\qnrel{S}}^\pr}{\qnjac{J}^\pr}{\mproj{\qnjac{J}}^\pr}
\clebsch{l}{ m_l}{\tfrac{1}{2}}{m_{s_3}}{j}{ m_{j}} 
\clebsch{l'}{m'_l}{\tfrac{1}{2}}{m'_{s_3}}{j'}{m'_{j}} 
\clebsch{\qnjac{J}}{\mproj{\qnjac{J}}}{j}{m_j}{\mathcal{J}}{M_{\mathcal{J}}} 
\clebsch{\qnjac{J}^\pr}{\mproj{\qnjac{J}}^\pr}{j'}{ m'_j}{\mathcal{J}}{ M_{\mathcal{J}}} 
\clebsch{\bar{T}}{M_{\bar{T}}}{\tfrac{1}{2}}{m_{t_3}}{\mathcal{T}}{M_{\mathcal{T}}}
\clebsch{\bar{T}'}{M_{\bar{T}}}{\tfrac{1}{2}}{m_{t_3}}{\mathcal{T}}{M_{\mathcal{T}}}
\left< p' q' \alpha' \left| V_{\text{3N}} \right| p q \alpha \right> \,,
\end{split}
\raisetag{6.5\baselineskip}
\end{align}
\end{widetext}
where we used the orthonormality of the spherical harmonics, that the two-body spin is given by $S=\bar{S}$ and $S'=\bar{S}'$, and $\mathbf{k}'_3 = \mathbf{P} + \mathbf{k}_3 - \mathbf{P}'$. A significant number of the sums in this expression can be reduced analytically, while several remaining sums and integrals need to be performed numerically. The final result can be expressed in the following simplified form:
\begin{align}
\label{eq:Gamma_Jacobi_short}
\begin{split}
& \bigl< p' P' \gamma' \bigl| \Gamma^{\text{3N}} \bigr| p P \gamma \bigr> \\
&= \sum_{\substack{\mathcal{J} \mathcal{T}\\l l' j j'}}
\int d\hat{\mathbf{P}} d \hat{\mathbf{P}}' \frac{d \mathbf{k}_3}{( 2\pi)^3} 
\left< p' q' \alpha' | V_{\text{3N}} | p q \alpha \right>  \\
&\quad \times \sum_{ \substack{j_x j_y j_z\\ m_y m_z}} \sum_{M_{\bar{L}_{\text{CM}}} M'_{\bar{L}_{\text{CM}}}} \sum_{m_l  m'_{l}} 
A^{\alpha \alpha' \gamma \gamma'}_{j_x j_y j_z} B^{\bar{T} \bar{T}' M_{\bar{T}} \mathcal{T}}_{j_x j_y j_z m_{y} m_z} (\mathbf{k}_3, \mathbf{k}'_3) \\
&\quad \times \clebsch{l}{m_l}{\bar{L}'_{\text{CM}}}{M'_{\bar{L}_{\text{CM}}}}{j_y}{m_{y}} \clebsch{l'}{m'_l}{\bar{L}_{\text{CM}}}{M_{\bar{L}_{\text{CM}}}}{j_z}{m_{z}} \\ & \quad \times Y^*_{\bar{L}_{\text{CM}} M_{\bar{L}_\text{CM}}} (\hat{\mathbf{P}})\, Y_{\bar{L}'_{\text{CM}} M'_{\bar{L}_\text{CM}}} (\hat{\mathbf{P}'})\, Y^*_{l m_l} (\hat{\mathbf{q}})\, Y_{l' m'_l} (\hat{\mathbf{q}}') \,,
\end{split}
\raisetag{4.5\baselineskip}
\end{align}
with the superscripts $\alpha$ and $\gamma$ indicating the dependence on all of the quantum numbers defined in Eqs.~\eqref{eq:alpha_basis} and~\eqref{eq:gamma_basis}, respectively.
We introduced the following quantities
\begin{align}
\begin{split}
A^{\alpha \alpha' \gamma \gamma'}_{j_x j_y j_z} &= \delta_{\qnrel{J} \qnjac{J}} \delta_{\qnrel{J^\pr} \qnjac{J^\pr}} (-1)^{-\qnrel{J} + \qnrel{J^\pr} + j + j' + 2 j_x + j_y - j_z}  \hat{\mathcal{J}}^2 \hat{j} \hat{j}' \hat{j_y} \hat{j_z} \hat{j}_x^2 \\ &\quad \times \sixj{\bar{L}'_{\text{CM}}}{l}{j_y}{\frac{1}{2}}{j_x}{j} 
\sixj{\bar{L}_{\text{CM}}}{l'}{j_z}{\frac{1}{2}}{j_x}{j'} \ninej{j_x}{\bar{L}'_{\text{CM}}}{j}{\bar{L}_{\text{CM}}}{\bar{J}_{\text{tot}}}{\bar{J}}{j'}{\bar{J}'}{\mathcal{J}} \,,
\end{split}
\raisetag{2.5\baselineskip}
\end{align}
and    
\begin{align}
\begin{split}
&B_{j_x j_y j_z m_y m_z}^{\bar{T} \bar{T}' M_{\bar{T}} \mathcal{T}} (\mathbf{k}_3, \mathbf{k}'_3) = \\
&\sum_{M_{\mathcal{T}}} \sum_{j_u m_u} \sum_{\substack{n_3 l_3 j_3 m_{j_3} \\m_{l_3} m'_{l_3} m_{t_3}}} \hat j_3^2 \clebsch{l_3}{m_{l_3}}{j_z}{m_z}{j_u}{m_u} \clebsch{l_3}{m'_{l_3}}{j_y}{m_y}{j_u}{m_u}  \:  \clebsch{\bar{T}}{M_{\bar{T}}}{\tfrac{1}{2}}{m_{t_3}}{\mathcal{T}}{M_{\mathcal{T}}} \clebsch{\bar{T}}{M_{\bar{T}}}{\tfrac{1}{2}}{m_{t_3}}{\mathcal{T}}{M_{\mathcal{T}}} \\
&  \times \ninej{j_x}{\frac{1}{2}}{j_y}{\frac{1}{2}}{j_3}{l_3}{j_z}{l_3}{j_u} Y_{l_3 m'_{l_3}} (\hat{\mathbf{k}}'_3)\, Y_{l_3 m_{l_3}}^* (\hat{\mathbf{k}}_3) \\
&\times \varphi_{n_3 l_3 j_3 m_{j_3} m_{t_3}} (k_3)\, \varphi_{n_3 l_3 j_3 m_{j_3} m_{t_3}} (k'_3) \,,
\end{split}
\raisetag{3.5\baselineskip}
\end{align} 
with $\hat{j} = \sqrt{2 j +1}$ for all angular momentum quantum numbers. In addition we introduced the auxiliary quantum numbers $j_x$, $j_y$, $j_z$, and $j_u$ as well as their projections $m_y$, $m_z$, and $m_u$. These intermediate quantities are obtained by coupling $l$ with $\bar{L}'_\text{CM}$ to $j_y$, $l'$ with $\bar{L}_\text{CM}$ to $j_z$, $l_3$ with $j_y$ and $j_z$ to $j_u$, and $s=1/2$ with $j_y$ and $j_z$ to $j_x$.
Note that the values of the
Jacobi momenta $q$ and $q'$ are implicitly fixed by the relations $\mathbf{k}_3 = 3/2 \mathbf{q} + \mathbf{P}/2$ and $\mathbf
{k}'_3 = \mathbf{P} + \mathbf{k}_3 - \mathbf{P}'$. Furthermore,
$A^{\alpha \alpha' \gamma \gamma'}_{j_x j_y j_z}$ does not depend on $m_{y}$, $m_z$, or any
of the momenta in the spherical harmonics, such that it can be  evaluated independently of the sums and integrals in Eq.~\eqref{eq:Gamma_Jacobi_short} and can be easily prestored. All other quantities that
involve those quantum numbers, like $M_{\bar{L}_{\text{CM}}}$ and $m_l$, that also enter the
spherical harmonics need to be recomputed for each point of the momentum mesh in the numerical integration.

In order to take into account all possible recoupling contributions from the Jacobi representation to a single-particle representation, matrix elements for sufficiently large values of the angular momentum quantum numbers in the basis need to be computed for a given single-particle basis size $e_{\text{max}}$. Generally, the matrix elements of $\Gamma^{\text{3N}}$ get systematically suppressed as the values of the total angular momentum quantum numbers $\bar{J}$, $\bar{L}_{\text{CM}}$, and $\bar{J}_{\text{tot}}$ increase.

The integrals and sums in Eq.~\eqref{eq:Gamma_Jacobi_short} are then calculated for a given partial-wave truncation specified by $\bar{L}_{\text{CM}}^{\text{max}}$ and $\bar{J}_{\text{tot}}^{\text{max}}$ in the partial-wave basis $\gamma$ for a specified number of mesh points of the four momenta $p$, $P$, $p'$, and $P'$. Our calculations show that for the interaction employed in this work about 20 points for each of these momenta are needed for calculations up to around $^{48}$Ca, while for heavier systems higher partial-wave channels become relevant which make it necessary to also increase the number of mesh points. Specifically, for our largest calculations of $^{132}$Sn and $^{208}$Pb we used $N_P = 35$ and $N_p = 20$. 

\subsection{Transformation to HO basis}
\label{sec:rel_HO_trafo}

Once the matrix elements of $\bigl< p' P' \gamma' \bigl| \Gamma^{\text{3N}} \bigr| p
P \gamma \bigr>$ have been computed in the plane-wave Jacobi representation, the next step consists of
transforming these relative momentum matrix elements to a
relative harmonic oscillator basis of the form $\ket{N N_{\text{CM}} \gamma }$, where $N$ and $N_{\text{CM}}$ are the radial oscillator quantum numbers that
correspond to the relative momentum $p$ and two-body CM momentum
$P$, respectively, and $\gamma$ is the collective index of Eq.~\eqref{eq:gamma_basis}.
This step is straightforward as it represents a simple
generalization of the standard transformation expressions for free-space NN interactions. We additionally have to include the CM dependence and can then represent the plane-wave Jacobi states in Eq.~\eqref{eq:Gamma_Jacobi_short} in the relative HO basis by
\begin{align}
    \ket{N N_\text{CM}  \gamma } =
    \int dp\, dP\, p^2 P^2 R_{N\bar{L}}(p,b_{\text{rel}}) R_{N_\text{CM} \bar{L}_\text{CM}}(P,b_{\text{CM}}) \ket{pP\gamma}
    \,,
    \label{eq:gamma_HO_basis}
\end{align}
with the oscillator lengths $b_i = 1/\sqrt{M \omega_i}$ and the radial part $R_{N\bar{L}}$ of the harmonic oscillator wave function in momentum space is given by $R_{N\bar{L}} (p,b) = \braket{p \bar{L} |N \bar{L} }$. The oscillator parameters are related by \cite{Hebe203NF}
\begin{equation}
    b_{\text{rel}} = \sqrt{2} \, b\, , \quad b_{\text{CM}} = \frac{1}{\sqrt{2}} \, b \, ,
\end{equation}
where $b$ is the single-particle oscillator length of the HO basis.

Note that the number of partial-wave channels is much higher in the present case than for free-space two-body interactions due to the dependence of the matrix elements on the $\bar{L}_{\text{CM}}$ quantum number. This allows for more complicated angular-momentum coupling patterns in the basis $\left| \gamma \right>$ compared to the free-space two-body basis. 

In the following, we characterize the basis space in the Jacobi representation by the
total energy quantum number $\bar{E}^{(2)}$, which involves relative and
CM quantum numbers:
\begin{equation}
\bar{E}^{(2)} = 2 N_{\text{CM}} + \bar{L}_{\text{CM}} + 2 N + \bar{L} = e_1 + e_2 \, .
\label{eq:E2max_def}
\end{equation}
This implies that for $\bar{E}^{
(2)}_{\text{max}} \ge 2 e_{\text{max}}$ the NO does not involve any truncations on radial HO quantum numbers in a given
single-particle basis size of $e_{\text{max}}$. An overview of the quantum numbers and truncation schemes in both NO frameworks is given in Table~\ref{tab:truncations}.

\begin{figure}[t!]
    \centering
    \includegraphics[width=\columnwidth,clip=]{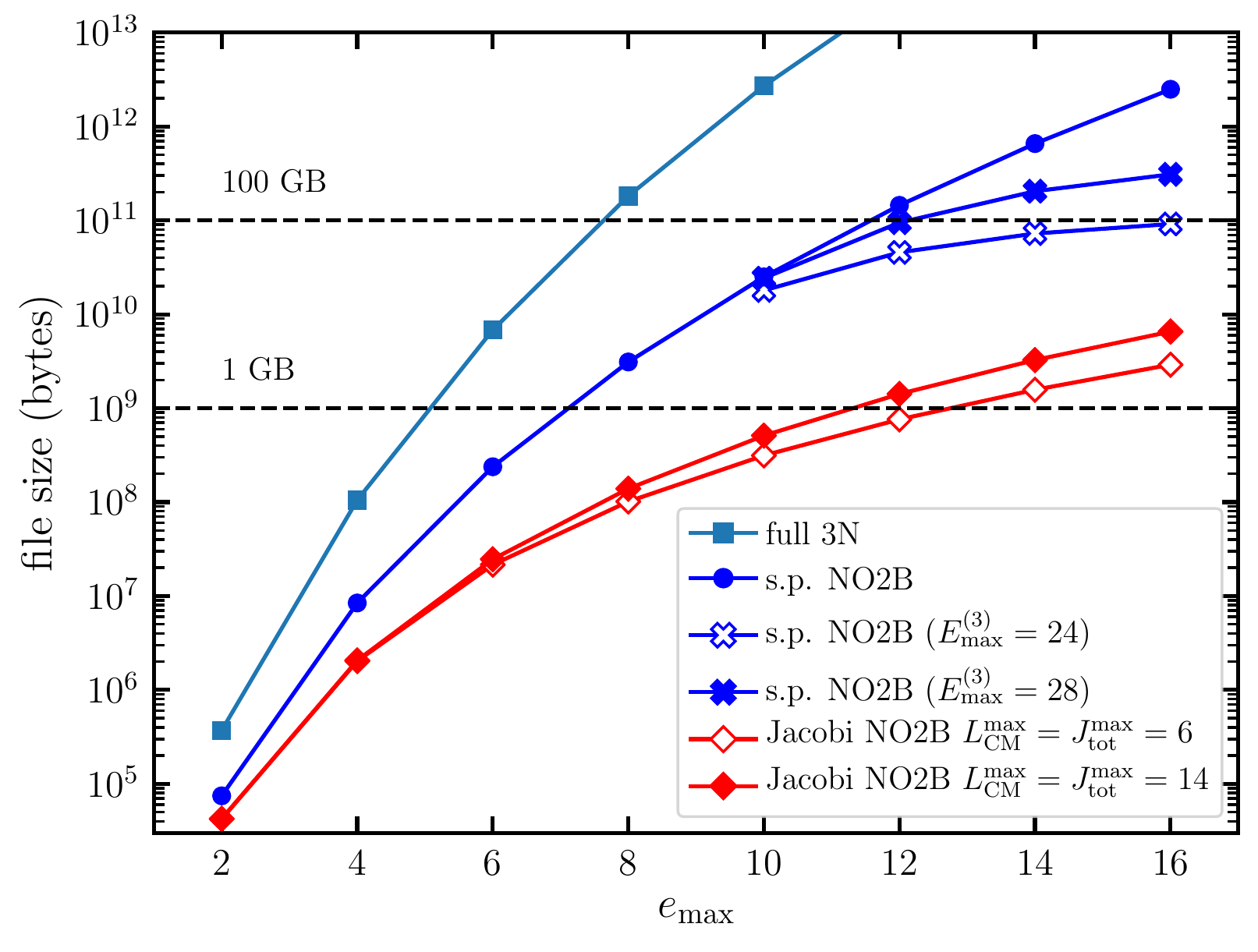}
    \caption{Memory requirements for the storage of intermediate operators in the different NO frameworks as a function of the single-particle basis size $e_{\text{max}}$. The light blue line refers to the dimension of the full three-body operator in a single-particle representation with $E^{(3)}_{\text{max}} = 3 e_{\text{max}}$. The dark blue lines show the storage space for only those 3N matrix elements needed in the NO2B approximation (see Ref.~\cite{Miya21E3max}), without $E^{(3)}_{\text{max}}$ cut (circles) and using the  cuts $E^{(3)}_{\text{max}}=24$ and $28$ (crosses). The red lines show the dimension of $\Gamma^{\text{3N}}$ in the basis defined in Eq.~(\ref{eq:gamma_HO_basis}) for $E^{(2)}_{\text{max}} = 2 e_{\text{max}}$ and the truncations $\bar{L}^{\text{max}}_{\text{CM}},\bar{J}^{\text{max}}_{\text{tot}}=6$ and $14$.}
    \label{fig:dims}
\end{figure}

In Fig.~\ref{fig:dims} we compare the memory requirements for storing the relevant intermediate matrix elements in the different NO frameworks. The light blue line shows the dimensions of the full 3N operator in a given single-particle basis size $e_{\text{max}}$ without employing any additional truncations, i.e., $E^{(3)}_{\text{max}} = 3 e_{\text{max}}$. In this case the matrix sizes become intractable very quickly and this option is consequently not viable for practical applications. When employing the new storage scheme of Ref.~\cite{Miya21E3max}, the dimensions get reduced significantly (dark blue lines), allowing to push the NO2B limits towards larger basis sizes, especially when applying additional cuts on $E^{(3)}_{\text{max}}$ (crosses). The displayed cuts $E^{(3)}_{\text{max}} = 24$ and $28$ provide sufficiently well converged calculations for soft interactions up to $^{132}$Sn and even $^{208}$Pb, as shown in Refs.~\cite{Miya21E3max,Hu21Pb208skin} and also in Sec.~\ref{sec:res} below. Finally, we show the storage space required for $\Gamma^{\text{3N}}$ in the HO basis defined in Eq.~(\ref{eq:gamma_HO_basis}) in the new Jacobi NO framework. Here we show the two cases $\bar{L}^{\text{max}}_{\text{CM}} = \bar{J}^{\text{max}}_{\text{tot}}=6$ and $14$. For both, the file sizes are significantly smaller than for the single-particle NO framework.

\begin{table}[t]
    \centering
    \renewcommand*{\arraystretch}{1.5}    
    \begin{tabularx}{\columnwidth}{l|l}
        \hline \hline
        \multicolumn{2}{l}{Angular-momentum quantum numbers} \\ \hline
        $\bar{L}_{\text{CM}} \, [\gamma]$ & two-body CM orbital angular momentum of $\mathbf{P}$ \\ \hline
        $\bar{J}_\text{tot} [\gamma]$ & total angular momentum of $\mathbf{p}$ and $\mathbf{P}$ \\ \hline  
        $\bar{L} \, [\gamma] = L \, [\alpha]$ & relative orbital angular momentum of $\mathbf{p}$\\ \hline
        $\bar{J} \, [\gamma] = J \, [\alpha]$ & total angular momentum of $\mathbf{p}$ \\ \hline
        $l \, [\alpha]$ & orbital angular momentum of $\mathbf{q}$ \\ \hline 
        $j \, [\alpha]$ & total angular momentum of $\mathbf{q}$ \\ \hline        
        $\mathcal{J} [\alpha]$ & total three-body angular momentum of $\mathbf{p}$ and $\mathbf{q}$ \\ \hline
        $l_i $ & single-particle orbital angular momentum of $\mathbf{k}_i$ \\
        \hline \hline
        \multicolumn{2}{l}{Energy quantum numbers} \\ \hline
        \multicolumn{2}{l}{$e_i = 2 n_i + l_i$} \\ 
        \multicolumn{2}{l}{$E^{(3)} = e_1 + e_2 + e_3$} \\
        \multicolumn{2}{l}{$\bar{E}^{(2)} [\gamma] = 2 N_{\text{CM}} + \bar{L}_{\text{CM}} + 2 N + \bar{L} = e_1 + e_2$} \\
        \hline \hline
        \multicolumn{2}{l}{General truncations} \\ \hline
        \multicolumn{2}{l}{$\bar{J} \, [\gamma] \le \bar{J}^{\text{max}}, \quad l \, [\alpha] \le l^{\text{max}}, \quad \mathcal{J} [\alpha] \le \mathcal{J}^{\text{max}}, \quad e_i \le e_{\text{max}}$} \\
        \hline \hline
        \multicolumn{2}{l}{Additional truncations in Jacobi normal ordering} \\ \hline
        \multicolumn{2}{l}{$\bar{L}_{\text{CM}} \, [\gamma] \le \bar{L}_{\text{CM}}^{\text{max}}$, \quad $\bar{J}_{\text{tot}} \, [\gamma] \le \bar{J}_{\text{tot}}^{\text{max}}$} \\
        \hline \hline
        \multicolumn{2}{l}{Additional truncations in single-particle normal ordering} \\ \hline
        \multicolumn{2}{l}{$E^{(3)} \le E^{(3)}_{\text{max}}$} \\
        \hline \hline
        \end{tabularx}    
    \caption{
    \label{tab:truncations}
    Summary of the relevant quantum numbers and employed truncations in the Jacobi and single-particle NO frameworks. The argument in square brackets indicates the basis in which the corresponding quantum number is defined, i.e., Eqs.~(\ref{eq:alpha_basis}) and (\ref{eq:gamma_basis}), respectively. Note that for $\bar{E}^{(2)}_{\text{max}} \ge 2 e_{\text{max}}$ there are no truncations on the radial HO quantum numbers in the Jacobi NO framework.}
\end{table}

\subsection{Transformation to single-particle basis}
\label{sec:sp_trafo}

For the discussion of the final transformation to single-particle states, we consider
$JT$-coupled two-body states in the basis defined in Eq.~\eqref
{eq:HO_single_particle_coupled}
\begin{align}
\left| (a b) \bar{J}_{\text{tot}} M_{\bar{J}_{\text{tot}}} \bar{T} M_{\bar{T}} \right> &= \sum_{\substack{m_{j_a} m_{j_b} \\ m_{t_a} m_{t_b}}} \mathcal{C}_{j_a m_{j_a} j_b m_{j_b}}^{\bar{J}_{\text{tot}} M_{\bar{J}_{\text{tot}}}} \mathcal{C}_{\tfrac{1}{2} m_{t_a} \tfrac{1}{2} m_{t_b}}^{\bar{T} M_{\bar{T}}} \left| a b \right> \,,
\end{align}
by coupling the single-particle angular momenta and isospins to the total angular momentum $\bar{J}_\text{tot}$ and $\bar{T}$, respectively,
with the uncoupled single-particle states given by
\begin{align}
    \left| a b \right> &= \left| n_a (l_a \tfrac{1}{2}) j_a m_{j_a} \tfrac{1}{2} m_{t_a} n_b (l_b \tfrac{1}{2}) j_b m_{j_b} \tfrac{1}{2} m_{t_b} \right> \, .
\end{align}

Many-body frameworks usually require as input $JT$-coupled matrix elements in the single-particle basis
\begin{align}
    \left< (a b) \bar{J}_{\text{tot}} \bar{T}' M_{\bar{T}} \right| \Gamma^{\text{3N}} \left| (c d) \bar{J}_{\text{tot}} \bar{T} M_{\bar{T}} \right> \, .
\end{align}
Note that in general $\Gamma^{\text{3N}}$ has nonvanishing matrix
elements for off-diagonal $\bar{T},\bar{T}'$ and will
depend on $M_{\bar{T}}$, while it is diagonal in $\bar{J}_\text{tot}$ and independent of $\mproj{\bar{J}_\text{tot}}$, as discussed for Eq.~\eqref{eq:gamma_basis}. However, since the transformation to the single-particle basis does not modify the isospin dependence of the states, we will in the following suppress the
isospin quantum numbers for simplicity. In the first step of the transformation,
we factorize the spin part of the states by recoupling the two-body
states from a $j$-coupling scheme to an $ls$-coupling scheme:
\begin{align}
    & \left| \left[ n_a(l_a s_a) j_a n_b(l_b s_b) j_b \right] \bar{J}_{\text{tot}} \right> \nonumber \\
    &= \sum_{\lambda S} \hat{j}_a \hat{j}_b \hat{\lambda} \hat{S} \ninej{l_a}{s_a}{j_a}{l_b}{s_b}{j_b}{\lambda}{S}{\bar{J}_{\text{tot}}} \left| \left[ (n_a l_a n_b l_b) \lambda (s_a s_b) S \right] \bar{J}_{\text{tot}} \right> \, ,
\end{align}
where the orbital angular momenta $l_a$ and $l_b$ couple to the total orbital angular momentum $\lambda$. 
In order to make use of the standard definitions of the Talmi-Moshinsky transformation brackets of Ref.~\cite{Kamu01TalmiMos}, which are given by
\begin{align}
 \left| \left( n_a l_a n_b l_b \right) \lambda \right> &= \sum_{N_{\text{CM}} \bar{L}_{\text{CM}} N \bar{L}} \left< (N_{\text{CM}} \bar{L}_{\text{CM}} N \bar{L}) \lambda \big| (n_a l_a n_b l_b) \lambda \right> \nonumber \\
 & \quad \times \left| \left( N_{\text{CM}} \bar{L}_{\text{CM}} N \bar{L} \right) \lambda \right> \,,
\end{align}
we need to recouple also the angular momenta of the Jacobi HO basis defined in Eq.~(\ref{eq:gamma_basis}) to first couple the relative and CM angular momentum to $\lambda$:
\begin{align}
\left| N_\text{CM} N \left[ ( \qncm{\bar{L}} \qnrel{L}) \lambda \qnrel{S}) \right] \qntot{\bar{J}} \right> &= \sum_{\bar{J}} (-1)^{\bar{L}+\bar{S}+\bar{J}} \hat{\bar{J}} \hat{\lambda} \sixj{\qnrel{S}}{\qnrel{L}}{\qnrel{J}}{\qncm{\bar{L}}}{\qntot{\bar{J}}}{\lambda} \notag \\
& \quad \times \left| N_\text{CM} N \left[ (\qnrel{L} \qnrel{S}) \qnrel{J} \qncm{\bar{L}} \right] \qntot{\bar{J}} \right> \,,
\end{align}
where we changed the coupling order of $\bar{L}$ and $\bar{L}_\text{CM}$, which introduces an additional phase.
Summarizing, this leads to the final transformation relation for the (non-normalized) two-body states:
\begin{widetext}
\begin{align}
    \left| (a b) \bar{J}_{\text{tot}} \right> 
    &= \sum_{\lambda \bar{S}} \hat{j}_a \hat{j}_b \hat{\lambda}^2 \hat{\bar{S}} \hat{\bar{J}} \ninej{l_a}{s_a}{j_a}{l_b}{s_b}{j_b}{\lambda}{\bar{S}}{\bar{J}_{\text{tot}}} \sum_{\substack{N_{\text{CM}} \bar{L}_{\text{CM}} \\ N \bar{L}}} \left< (N_{\text{CM}} \bar{L}_{\text{CM}} N \bar{L}) \lambda \big| (n_a l_a n_b l_b) \lambda \right> \sum_{\bar{J}} (-1)^{\bar{L} + \bar{S} + \bar{J}} \sixj{\qnrel{S}}{\qnrel{L}}{\qnrel{J}}{\qncm{\bar{L}}}{\qntot{\bar{J}}}{\lambda} \left| N_{\text{CM}} N \left[ (\bar{L} \bar{S}) \bar{J} \bar{L}_{\text{CM}} \right] \bar{J}_{\text{tot}} \right> \, .
\end{align}
\end{widetext}

\begin{figure*}[t]
\centering
\includegraphics[width=\textwidth,clip=]{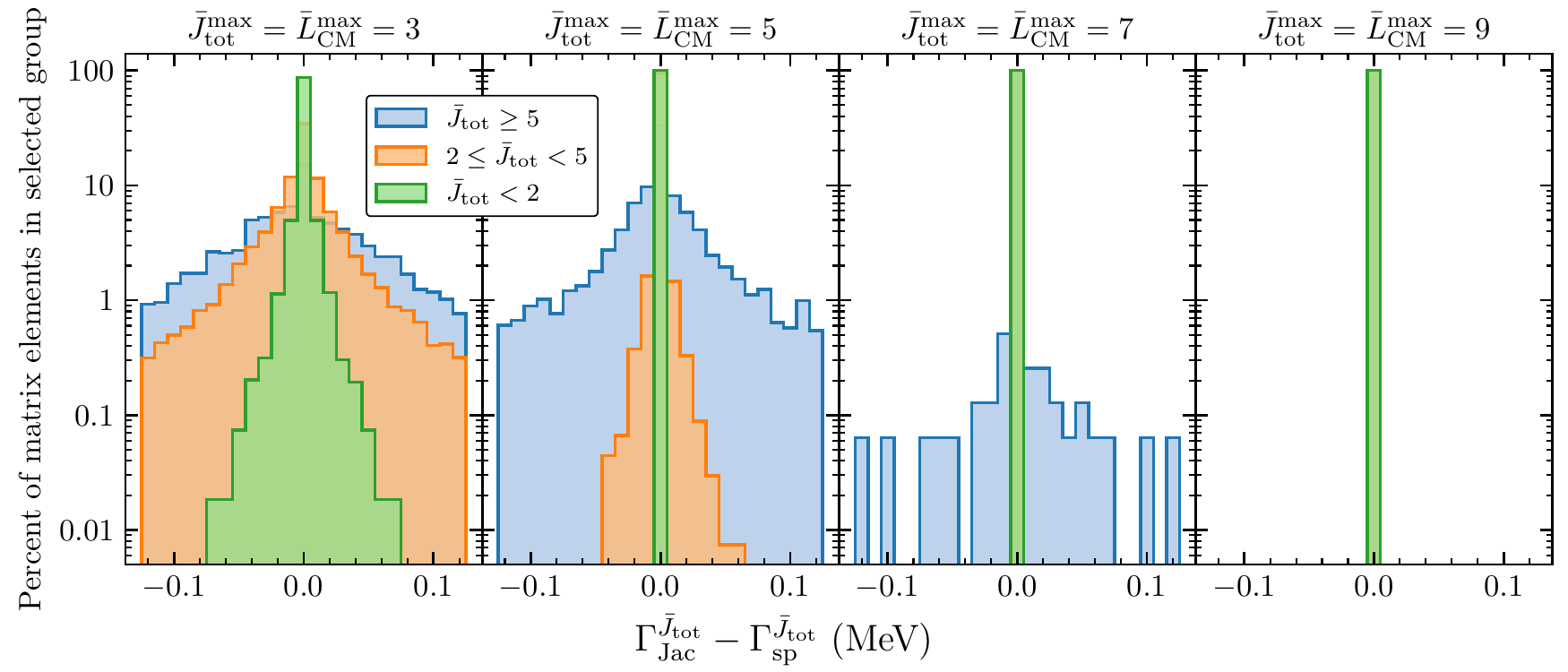}
\caption{
\label{fig:ME_comparison}
Distributions of differences $\Gamma_{\text{Jac}}^\text{3N} - \Gamma_{\text{sp}}^\text{3N}$ of two-body matrix elements in the Jacobi and single-particle NO for different $\bar{J}_\text{tot}$ blocks of $\bar{J}_\text{tot}<2$ (green), $2 \leq \bar{J}_\text{tot} \leq 5$ (orange), and $\bar{J}_\text{tot} \geq 5$ (blue) in the effective two-body interaction.
We show results for different truncations in the Jacobi NO, using $\bar{J}^\text{max}_\text{tot} = \bar{L}^\text{max}_\text{CM} = 3$, 5, 7, and 9  in the first through fourth panel, respectively. 
The y-axis shows the percentage of matrix elements with the difference specified on the x-axis in the selected $\bar{J}_\text{tot}$ block.
Results are shown for the 1.8/2.0 EM 3N interaction with $e_\text{max} = 4$, $E^{(3)}_\text{max}=12$, and $\hbar \omega = 16$~MeV using an $^{16}$O HF reference state.
}
\end{figure*}

\section{Results}
\label{sec:res}

\subsection{Matrix-element comparison}
\label{sec:ME_comparison}

Throughout this work, we use the chiral NN+3N interaction constructed in Ref.~\cite{Hebe11fits}, labeled as 1.8/2.0 EM. This interaction provides a good reproduction of ground-state energies over a large range of mass numbers from $A=4-132$~\cite{Simo17SatFinNuc,Morr17Tin,Stro21atomicNucl,Miya21E3max} as well as reasonable saturation properties of nuclear matter~\cite{Hebe11fits,Dris17MCshort} and hence offers an ideal test case for investigating the Jacobi NO framework.

We first benchmark the new NO framework by explicitly comparing matrix elements of the effective two-body interaction $\Gamma^\text{3N}$ in the single-particle and Jacobi approach. 
The reference state employed for all NO applications in the Jacobi basis in this work is given by an HF reference state calculated in the single-particle basis using the 1.8/2.0 EM interaction. The corresponding model space is characterized by the truncation of the single-particle states $e_\text{max}$ and the additional truncation in the space of three-body states $E^{(3)}_\text{max}$.
The HF orbitals can then be easily applied in the Jacobi NO based on Eq.~\eqref{eq:HF_ref_state}.
We note that the construction of the reference state still involves the representation of the 3N interaction in a single-particle basis. However, the HF solution is known to converge rapidly with respect to the model-space size. Consequently, values of $E^{(3)}_\text{max} \lesssim 16$ are typically sufficient to obtain sufficiently converged reference states for light to medium-mass nuclei. 
The reference-state dependence is further investigated in Sec.~\ref{sec:application_heavy}.

\begin{figure}[t]
\centering
\includegraphics[width=\columnwidth,clip=]{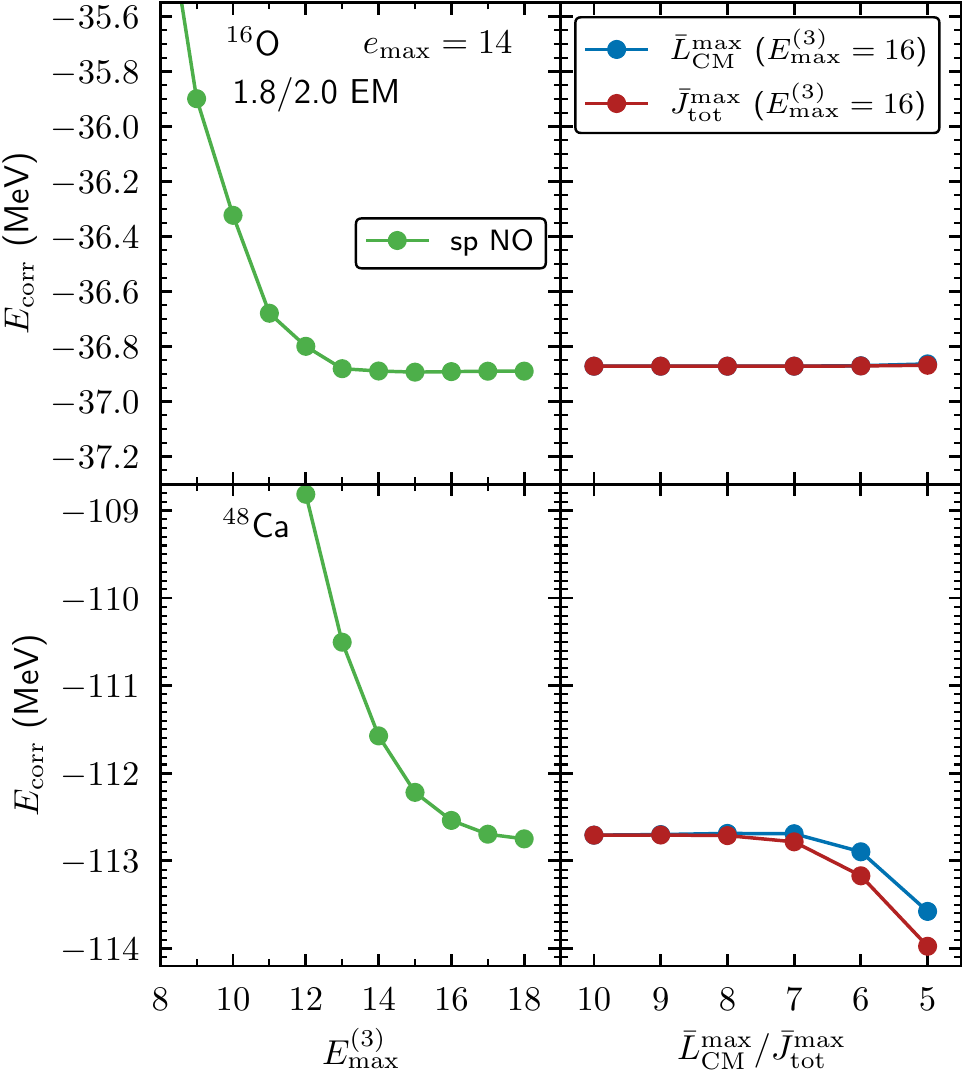}
\caption{
\label{fig:Ecorr_O16_Ca48_truncations}
Correlation energy in the single-particle (left panels) and Jacobi (right panels) NO frameworks as a function of $E^{(3)}_\text{max}$ (left panels) and $\bar{L}_\text{CM}^\text{max}$ or $\bar{J}_\text{tot}^\text{max}$ (right panels). In the right panels, the blue lines show $E_{\text{corr}}$ as a function of $\bar{L}_\text{CM}^\text{max}$ while keeping $\bar{J}_\text{tot}^\text{max}$ at its maximal value of 10, and the red lines show $E_{\text{corr}}$ as a function of $\bar{J}_\text{tot}^\text{max}$ while keeping $\bar{L}_\text{CM}^\text{max}$ at its maximal value of 10. We show results for $^{16}$O and $^{48}$Ca in the top and bottom panels, respectively, using the 1.8/2.0 EM interaction, an $e_\text{max}=14$ model space, and an oscillator frequency of $\hbar \omega = 16$~MeV in the HF basis. The Jacobi NO uses an $E^{(3)}_\text{max}=16$ HF reference state.}
\end{figure}

In Fig.~\ref{fig:ME_comparison}, we show the difference between the normal-ordered two-body matrix elements $\Gamma^\text{3N}$, defined in Eq.~(\ref{eq:NO_contraction}), in the Jacobi and single-particle basis for an $^{16}$O HF reference state.
We employ a model space of $e_\text{max} = 4$ with $E^{(3)}_\text{max}=12$ such that there is no active $E^{(3)}_\text{max}$ cut. The maximum total three-body angular momentum is set to $\mathcal{J}_{\text{max}}=9/2$, which is typically enough to obtain reasonably converged results for finite nuclei and nuclear matter~\cite{Hebe203NF}. 
Using  the same HF reference state in both frameworks allows for a clean comparison of the normal-ordered matrix elements.
For a better understanding of the different contributions to the two-body interaction, we divide the set of matrix elements into three blocks of total angular momentum $\bar{J}_\text{tot}$ in the two-body basis. We consider blocks of $\bar{J}_\text{tot} < 2$, $2 \leq \bar{J}_\text{tot} \leq 5$, and $\bar{J}_\text{tot} \geq 5$ and show the percentage of matrix elements as a function of the  absolute  difference in the corresponding blocks. By varying the $\bar{J}_\text{tot}^\text{max}$ and $\bar{L}_\text{CM}^\text{max}$ truncations in the Jacobi framework from 3 to 9, we see that increasing $\bar{J}_\text{tot}$ and $\bar{L}_\text{CM}$ systematically brings the matrix elements in the Jacobi in better agreement with the single-particle NO framework. In general, both truncations of $\bar{J}_\text{tot}$ and $\bar{L}_\text{CM}$ can be varied independently. However, here we truncate both quantum numbers at the same value for the sake of simplicity. Smaller $\bar{J}_\text{tot}$ blocks (green) already show an excellent agreement for small $\bar{J}_\text{tot}^\text{max}$ and $\bar{L}_\text{CM}^\text{max}$, while matrix elements for larger $\bar{J}_\text{tot}$ values (orange and blue) require larger values for a good agreement. For the highest truncation of $\bar{J}_\text{tot}^\text{max} = \bar{L}_\text{CM}^\text{max}=9$ shown here, both methods yield essentially identical matrix elements. This can be understood by the generalized transformation to the single-particle basis in Sec.~\ref{sec:sp_trafo} for the Jacobi framework. The two-body CM quantum number $\bar{L}_\text{CM}$ couples with $\bar{L}$ and $\bar{S}$ to the total angular momentum $\bar{J}_\text{tot}$, such that for a complete basis we would need to take $\bar{L}_\text{CM}^\text{max}$ to $\bar{J}_\text{tot}^\text{max} + \bar{L}^\text{max} + \bar{S}^\text{max}$, with $\bar{J}_\text{tot}^\text{max} = 9$ in an $e_\text{max}=4$ model space and $\bar{S}^\text{max}=1$. However, the contributions from very high $\bar{L}_\text{CM}$ in larger model spaces are found to be very small and the rightmost panel in Fig.~\ref{fig:ME_comparison} shows almost perfect agreement for $\bar{J}_\text{tot}^\text{max}=\bar{L}_\text{CM}^\text{max}=9$ for all $\bar{J}_\text{tot}$ blocks.

\subsection{Benchmarks for light and medium-mass nuclei}
\label{sec:truncation_benchmark_light}

We extend our benchmarks from a detailed comparison of matrix elements to ground-state energies of light and medium-mass nuclei, again comparing the Jacobi and single-particle NO frameworks.
All many-body calculations in this work use the \textit{ab initio} in-medium similarity renormalization group (IMSRG)~\cite{Herg16PR} with the publicly available IMSRG(2) solver by Stroberg~\cite{Stro17imsrggit}, and we denote the calculated ground-state energies by $E_\text{IMSRG}$. 
In the following we compare correlation energies defined by
\begin{align}
    E_\text{corr} = E_\text{IMSRG} - E_\text{HF} \,,
\end{align}
where $E_\text{HF}$ is the energy of the HF reference state.
We focus on this, because the HF solution converges faster than the many-body solution with respect to $E^{(3)}_\text{max}$, so that any residual energy dependence on $E^{(3)}_\text{max}$ is mostly due to correlation effects in the many-body expansion. 
For all following calculations, unless otherwise specified, we use a model-space size of $e_\text{max}=14$, which is sufficient to obtain converged results up to medium-mass nuclei~\cite{Simo17SatFinNuc}. In our calculations based on the 1.8/2.0 EM interaction we find that $e_\text{max}=14$ is sufficient for converged results up to $^{132}$Sn, only requiring $e_\text{max}=16$ for $^{208}$Pb. 
We emphasize that the two frameworks apply slightly different truncations regarding the relative angular momenta $J$ and $l$ in the 3N basis (see also Table~\ref{tab:truncations}). The Jacobi framework uses $J\le5$ and $l\le5$ for all three-body partial waves, whereas the single-particle approach uses a truncation of $J=8$, $J=7$, and $J=6$ for channels with total three-body angular momenta of $\mathcal{J} \leq 5/2$, $\mathcal{J}=7/2$, and $\mathcal{J} \ge 9/2$, respectively, while applying no explicit cuts on the $l$ values. The different choices are related to the fact that the Jacobi NO framework is based on antisymmetrized 3N matrix elements in momentum space, while the antisymmetrization for the traditional NO framework is typically performed in a complete HO subspace. Since the antisymmetrization in momentum space formally requires a complete sum over intermediate partial-wave quantum numbers (see also \cite{Hebe12msSRG,Hebe203NF}), it is advantageous to limit the number of partial-wave states to ensure proper antisymmetry of the retained states. However, the impact of matrix elements beyond $J=5$ for calculations of finite nuclei and nuclear matter is observed to be typically small~\cite{Hebe203NF}. Nevertheless, for our detailed benchmarks for heavier systems at the level of one per mille of the total binding energy we start to become sensitive to these details (see next section).

In Fig.~\ref{fig:Ecorr_O16_Ca48_truncations}, we show the correlation energy obtained within the traditional single-particle NO framework as a function of $E^{(3)}_\text{max}$ and in the Jacobi NO as a function of $\bar{L}_\text{CM}^\text{max}$ and $\bar{J}_\text{tot}^\text{max}$ for $^{16}$O and $^{48}$Ca.
For $E^{(3)}_\text{max}\approx16$ the effect of this truncation for light and medium-mass nuclei in the single-particle NO is known to be small. Here we show results up to 
$E^{(3)}_\text{max}=18$ in the left panels of Fig.~\ref{fig:Ecorr_O16_Ca48_truncations}, and use an $E^{(3)}_\text{max}=16$ reference state for the Jacobi NO in the right panels. Note that for the choices $E^{(3)}_\text{max}\leq18$ and $e_\text{max}=14$, we have an active cut in the single-particle NO (only $E^{(3)}_\text{max}\ge42$ would be complete), in contrast to the matrix-element comparison for small model spaces shown in Fig.~\ref{fig:ME_comparison}.

Converged correlation (and ground-state) energies in the single-particle approach are observed  around $E^{(3)}_\text{max}\approx 13$ and $E^{(3)}_\text{max}\approx18$ for $^{16}$O and $^{48}$Ca in the top and bottom panels of Fig.~\ref{fig:Ecorr_O16_Ca48_truncations}, respectively. The HF energy is already converged for smaller truncations of $E^{(3)}_\text{max}=10$ for $^{16}$O and $E^{(3)}_\text{max}=12$ for $^{48}$Ca (not shown).
In the Jacobi NO, we find that truncating the partial-wave quantum numbers at 
$\bar{L}_\text{CM}^\text{max}=\bar{J}_\text{tot}^\text{max} \approx 5$ is sufficient to obtain converged energies for $^{16}$O, while for $^{48}$Ca we need $\bar{L}_\text{CM}^\text{max}=\bar{J}_\text{tot}^\text{max} \approx 8$. This slightly larger truncation for $^{48}$Ca indicates that an increased number of partial-wave channels is required for converged results for larger mass numbers. We study this trend in more detail in the next section where we investigate heavier nuclei.
Comparing the results in the Jacobi and single-particle NO, we observe essentially perfect agreement of converged energies in both frameworks.
The only remaining dependence on $E^{(3)}_\text{max}$ in the Jacobi NO framework is given by the
reference-state calculation in the single-particle basis.
While we observe no residual dependence on the $E^{(3)}_\text{max}$ cut for the ground-state energies of $^{16}$O and $^{48}$Ca when using an $E^{(3)}_\text{max}=16$ reference state, 
the dependence on the reference state could become relevant when computing heavier nuclei. We also investigate this in the following section.

\begin{figure}[t]
\centering
\centering
\includegraphics[width=\columnwidth,clip=]{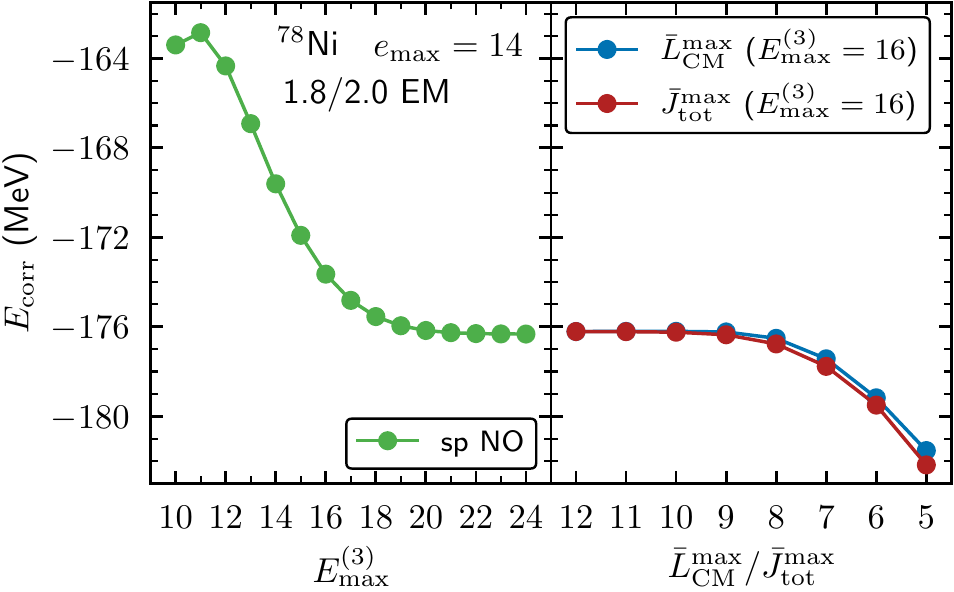}
\caption{
\label{fig:Ecorr_Ni78_truncations}
Same as Fig.~\ref{fig:Ecorr_O16_Ca48_truncations} but for $^{78}$Ni. 
}
\end{figure}

\subsection{Application to heavier systems}
\label{sec:application_heavy}

\begin{figure}[t]
\centering
\centering
\includegraphics[width=\columnwidth,clip=]{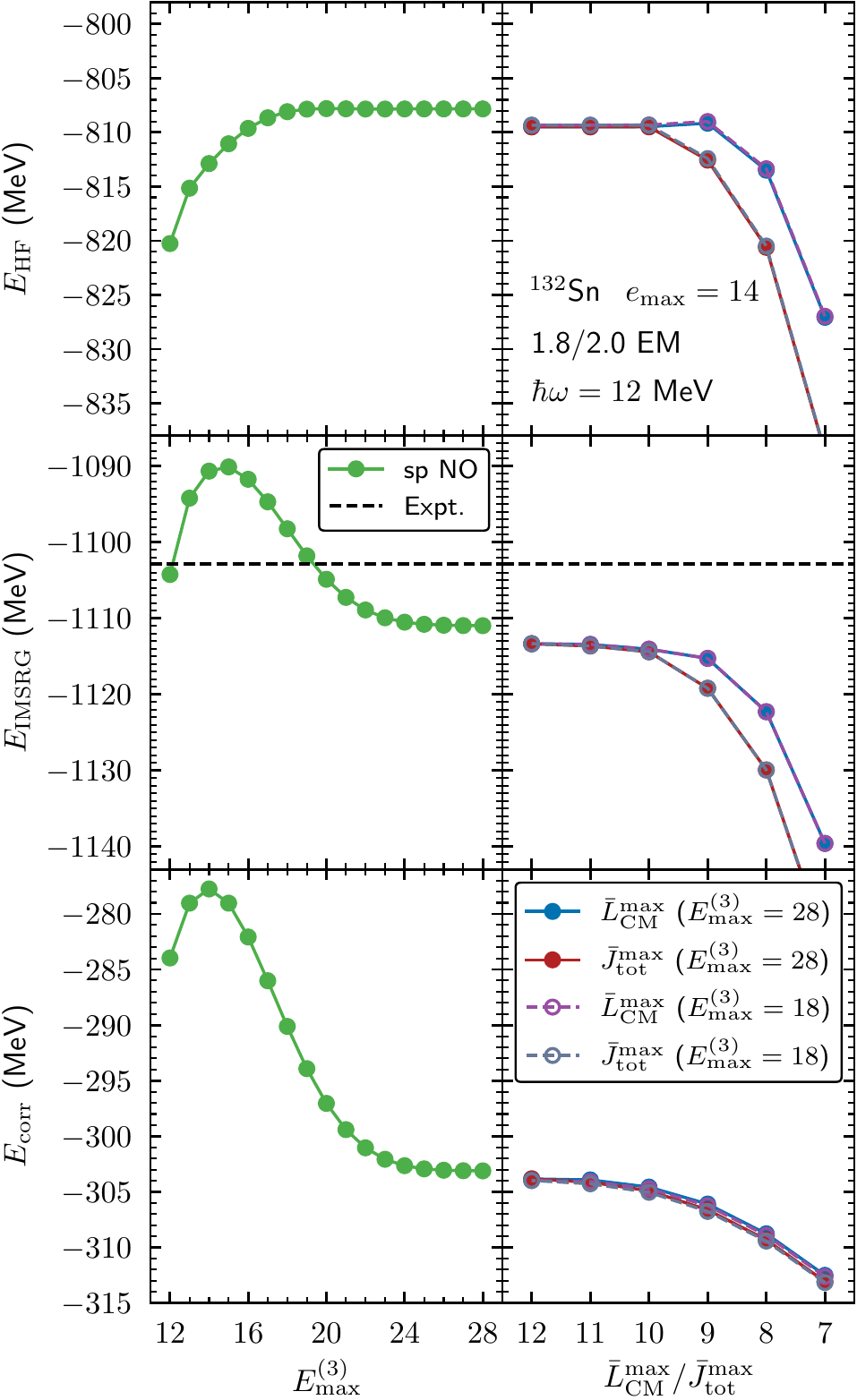}
\caption{
\label{fig:Ecorr_Sn132_EHF_Egs_Ecorr_hw12}
Same as Fig.~\ref{fig:Ecorr_Ni78_truncations} but for $^{132}$Sn and $\hbar \omega = 12$~MeV.
The Jacobi NO (right panels) uses an $E^{(3)}_\text{max}=28$ HF reference state (filled circles) and an $E^{(3)}_\text{max}=18$ HF reference state (open circles). We also show the HF energy and IMSRG energy in the top and middle panel, respectively, and increased the $E^{(3)}_\text{max}$ range in the single-particle NO (left) to $E^{(3)}_\text{max}=28$.
}
\end{figure}

We now turn our attention to heavier nuclei, first exploring the correlation energy of $^{78}$Ni in Fig.~\ref{fig:Ecorr_Ni78_truncations}. 
As discussed before, increasing the $E^{(3)}_\text{max}$ cut becomes important to obtain converged ground-state energies in the single-particle NO when approaching heavier systems. This can be seen in the left panel of Fig.~\ref{fig:Ecorr_Ni78_truncations}, where we study the correlation energy up to $E^{(3)}_\text{max}=24$ and find converged results for $E^{(3)}_\text{max} \approx 20$. Increasing the $E^{(3)}_\text{max}$ cut to higher values was possible due to Ref.~\cite{Miya21E3max}, and 3N matrix elements in the single-particle basis for the results in this section were provided by T.~Miyagi~\cite{Miya22private}. In contrast, for the Jacobi NO framework a basis size of $E^{(3)}_\text{max}=16$ for the reference state is already sufficient for converged IMSRG results. This can be understood based on the fact that HF calculations converge significantly faster than the IMSRG calculations. In fact, we find converged HF energies already around $E^{(3)}_\text{max} \approx 14$ for $^{78}$Ni (not shown). Consequently, no large-scale reference state calculations are needed in the Jacobi NO.

As for $^{16}$O and $^{48}$Ca, the Jacobi and single-particle NO frameworks provide the same converged energies. We note that by going to heavier nuclei, also the truncations $\bar{L}_\text{CM}^\text{max}$ and $\bar{J}_\text{tot}^\text{max}$ in the Jacobi NO have to be increased and we observe converged results for $\bar{L}_\text{CM}^\text{max}=\bar{J}_\text{tot}^\text{max} \approx 9$, slightly larger than what was observed for $^{48}$Ca.

\begin{table}[t]
    \centering
    \renewcommand*{\arraystretch}{1.5}    
    \begin{tabularx}{\columnwidth}{l|c|c}
          \hline\hline
          & $E_\text{HF}$ (MeV) & $E_\text{IMSRG}$ (MeV)\\ \hline 
          \multicolumn{3}{l} {Antisymmetrization in Jacobi HO basis}\\ \hline
          single precision & -806.11 & -1109.02 \\ 
          $J^{\text{max}}=l^{\text{max}}=5$ truncation & -808.79 & -1111.83  \\ 
          half precision & -807.84 & -1110.49\\
          \hline\hline
          \multicolumn{3}{l} {Antisymmetrization in Jacobi momentum-space basis}\\ \hline  
          single precision & -807.19 & -1110.27 \\  
          $J^{\text{max}}=l^{\text{max}}=5$ truncation & -809.05 & -1112.29 \\
          \hline\hline
          \multicolumn{3}{l} {Jacobi normal ordering}\\ \hline
          $\bar{L}_{\text{CM}}^{\text{max}} = \bar{J}_{\text{tot}}^{\text{max}} = 13$ & -809.49 & -1113.33 \\
          \hline\hline
    \end{tabularx}    
    \caption{
    \label{tab:spNO_uncertainties}
    Hartree-Fock and ground-state energies for $^{132}$Sn for an $e_\text{max}=14$ and $E^{(3)}_\text{max}=24$ basis space. Results are given for different 3N interaction files in the single-particle NO. The first three rows apply the antisymmetrization in the Jacobi HO basis, whereas the next two rows perform the antisymmetrization in the Jacobi momentum-space basis in comparison to the Jacobi NO results in the bottom row. If not stated otherwise, the single-particle NO uses the truncation of $J^{\text{max}}=8$, 7, and 6 (as described in the main text) and no $l$ truncation. The truncation $J^{\text{max}}=l^{\text{max}}=5$ is used in the Jacobi NO framework.}
\end{table}

The trends observed for $^{78}$Ni hold for even heavier systems. In Fig.~\ref{fig:Ecorr_Sn132_EHF_Egs_Ecorr_hw12}, we show a detailed comparison of the HF, IMSRG, and correlation energies of $^{132}$Sn in the single-particle and Jacobi NO.
The results in the single-particle basis are again shown up to $E^{(3)}_\text{max}=28$. Note that we employ an oscillator frequency of $\hbar \omega = 12$~MeV and include all matrix elements up to a total three-body angular momentum of $\mathcal{J}^\text{max}=15/2$ in order to be consistent with the choices of Ref.~\cite{Miya21E3max}.
As shown in the top panel of Fig.~\ref{fig:Ecorr_Sn132_EHF_Egs_Ecorr_hw12}, the HF energy converges rapidly with respect to $E^{(3)}_\text{max}$. This suggests that a reference state with $E^{(3)}_\text{max} \approx 18$, for which the HF calculation is already reasonably converged, is sufficient to obtain converged ground-state and correlation energies in the Jacobi NO framework. In fact, for the Jacobi NO we find essentially identical results using $E^{(3)}_\text{max}=28$ or $E^{(3)}_\text{max}=18$ for the reference state. The ground-state and correlation energy in the single-particle NO still depend on the $E^{(3)}_\text{max}$ truncation beyond $E^{(3)}_\text{max}=18$  and increasing this to $E^{(3)}_\text{max} \approx 24$ are required to obtain converged results. 

In the Jacobi NO, the $\bar{L}_\text{CM}^\text{max}$ and $\bar{J}_\text{tot}^\text{max}$ truncations required for converged calculations follow the trends observed for $^{78}$Ni in Fig.~\ref{fig:Ecorr_Ni78_truncations} and slightly increase to 
$\bar{L}_\text{CM}^\text{max} = \bar{J}_\text{tot}^\text{max} \approx 11$. The converged results for the ground-state energies of the two NO frameworks differ by about $2$ MeV. These deviations can be traced back to the slightly different angular momentum quantum number truncations, the different antisymmetrization methods for the 3N interaction matrix elements (see discussion in Sec.~\ref{sec:truncation_benchmark_light}), and the floating point precision employed in the two frameworks (in our work single precision or better is used, whereas Ref.~\cite{Miya21E3max} uses a combination of single and half precision). Table~\ref{tab:spNO_uncertainties} shows in detail the effects of the different choices for $^{132}$Sn.  First, it is obvious that even though the sensitivity of the results to these choices is rather small, at the few per-mille level, the effects are still noticeable in heavy nuclei. Second, the results obtained in the single-particle NO framework systematically approach the Jacobi NO results as the computational treatment of the interactions and the basis space size are adjusted properly. In particular, when using the exact same partial-wave truncations and antisymmetrization method of the 3N interaction in both frameworks, differences of only 500 keV and about 1 MeV remain for $E_\text{{HF}}$ and $E_{\text{IMSRG}}$, respectively (last two rows in the table).

\begin{figure}[t]
\centering
\centering
\includegraphics[width=0.95\columnwidth,clip=]{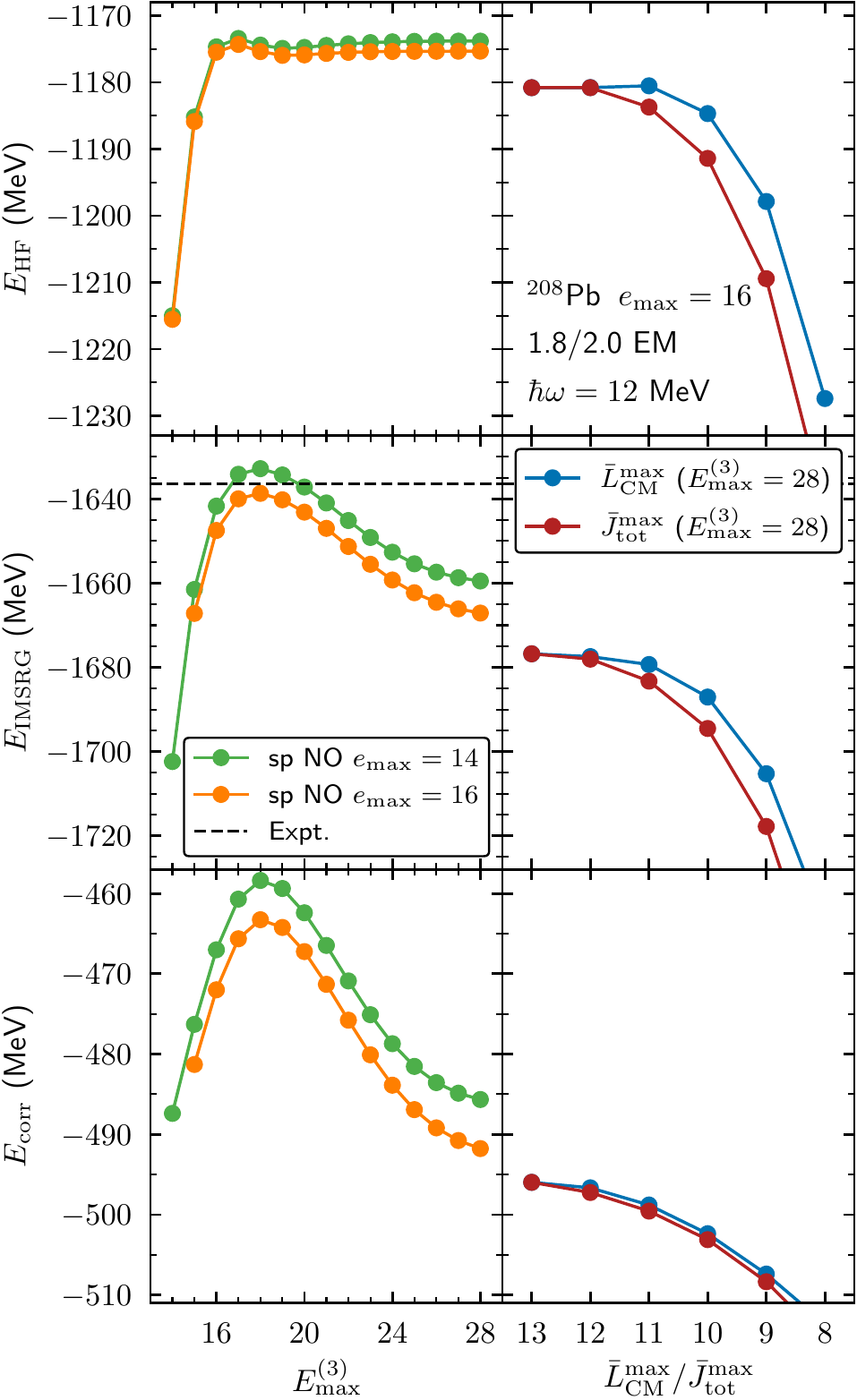}
\caption{
\label{fig:Ecorr_Pb208_EHF_Egs_Ecorr_hw12}
Same as Fig.~\ref{fig:Ecorr_Sn132_EHF_Egs_Ecorr_hw12} but for $^{208}$Pb. The single-particle NO results (left panels) are shown for $e_\text{max}=14$ and $e_\text{max}=16$ model spaces. The Jacobi NO framework (right panels) uses an $e_\text{max}=16$ and $E^{(3)}_\text{max}=28$ HF reference state.
}
\end{figure}

We further apply the Jacobi NO framework to even heavier systems, studying $^{208}$Pb in Fig.~\ref{fig:Ecorr_Pb208_EHF_Egs_Ecorr_hw12}.
For comparison we show calculations using an increased model space of $e_\text{max}=16$ in the single-particle approach.
As before, the HF energy is well converged around $E^{(3)}_\text{max}\approx20$. This is in contrast to the IMSRG energy, where the ground-state and thus the correlation energy are not fully converged with respect to $e_\text{max}$ and still show an $E^{(3)}_\text{max}$-sensitivity beyond $E^{(3)}_\text{max}=28$. 
The Jacobi framework shows similar trends as for $^{132}$Sn in Fig.~\ref{fig:Ecorr_Sn132_EHF_Egs_Ecorr_hw12}. We observe more binding for the HF and IMSRG energy compared to the single-particle NO and only slight differences for the correlation energy. As for $^{132}\text{Sn}$, the differences are again due to different choices of angular momentum quantum number truncations, antisymmetrization and floating point precision (see Table~\ref{tab:spNO_uncertainties}). Converged results to within 2~MeV (based on the convergence in $\bar{L}_\text{CM}^\text{max}$ and $\bar{J}_\text{tot}^\text{max}$) are obtained for $\bar{L}_\text{CM}^\text{max} = \bar{J}_\text{tot}^\text{max} \approx 13$, being only slightly larger than in $^{132}$Sn. 
We emphasize again that the reference-state construction is computationally cheap and that the final $e_\text{max}$ truncation for the effective two-body interaction can be scaled up in the Jacobi NO framework.

\begin{figure}
    \centering
    \includegraphics[width=0.95\columnwidth,clip=]{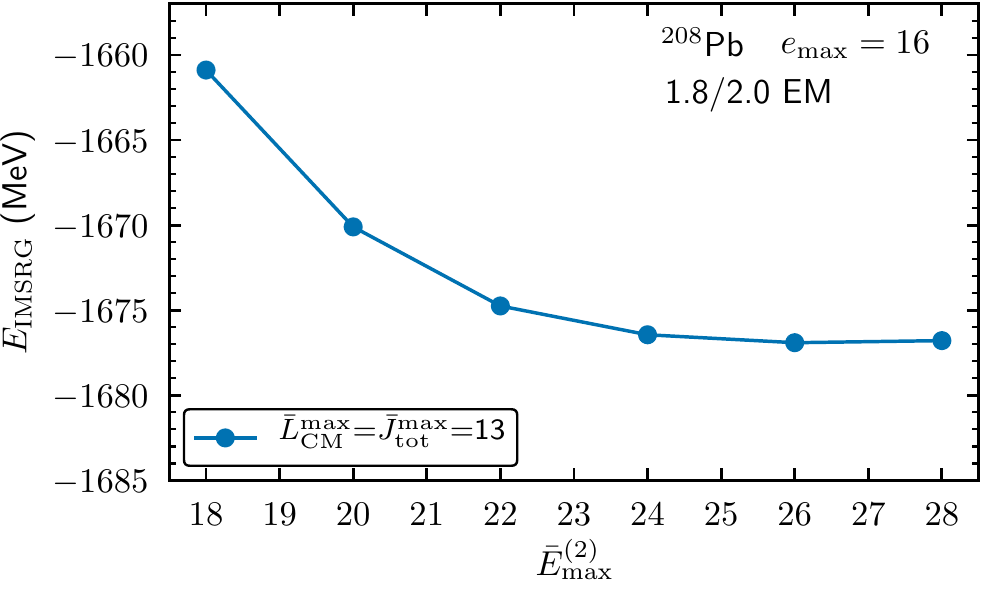}
    \caption{\label{fig:E2max_convergence}
    Ground-state energy energy of $^{208}$Pb based on the normal-ordered matrix elements obtained in the Jacobi NO framework as a function of the model-space parameter $E^{(2)}_{\text{max}}$ as defined in Eq.~(\ref{eq:E2max_def}). As in Fig.~\ref{fig:Ecorr_Pb208_EHF_Egs_Ecorr_hw12} an oscillator frequency of $\hbar \omega = 12$~MeV was used.
    }
\end{figure}

In Fig.~\ref{fig:E2max_convergence} we show the convergence behavior of the Jacobi NO results for the ground-state energy of $^{208}$Pb as a function of the $E^{(2)}_{\text{max}}$ truncation parameter using a single-particle basis size of $e_{\text{max}}=16$. In this case, $E^{(2)}_{\text{max}} \ge 32$ corresponds to untruncated NO calculations in the radial HO quantum numbers, while smaller values imply some cuts on $N_{\text{CM}}$ and $N$ (see Eq.~(\ref{eq:E2max_def}) and the related discussion). The results of the figure clearly show that $E^{(2)}_{\text{max}} = 26$ is already sufficient for obtaining practically converged calculations for $^{208}$Pb.

All these results highlight the versatility of the novel Jacobi NO framework, being able to target heavy nuclei in good agreement with standard NO methods and without being limited by an $E^{(3)}_\text{max}$ truncation. The only remaining dependence on the single-particle basis, due to the reference-state construction, is found to be resolved once the HF solution is converged. 

\subsection{Charge radii}
\label{subsec:JacNO_charge_radii}

\begin{figure}[t]
\centering
\includegraphics[width=\columnwidth,clip=]{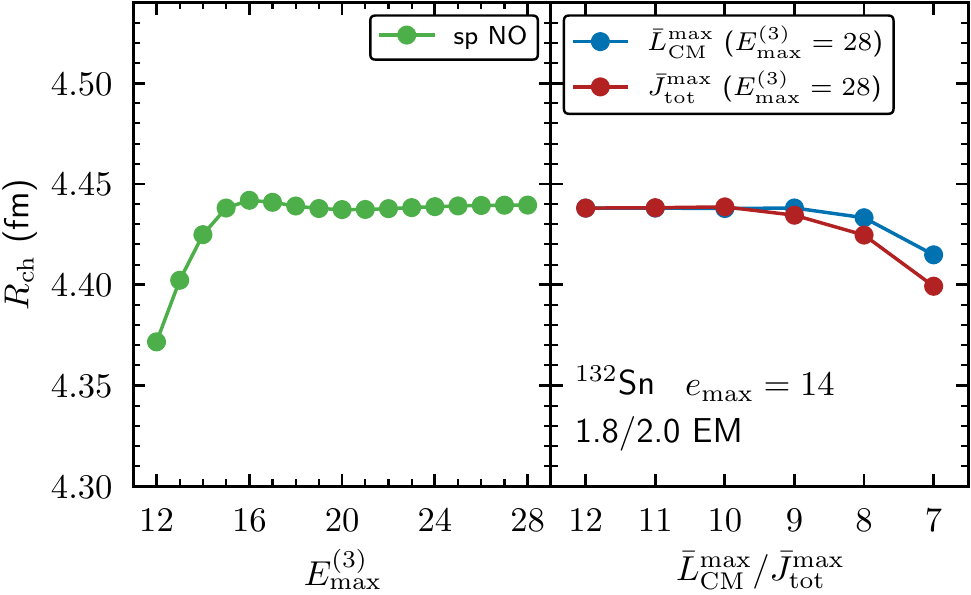}
\caption{
\label{fig:Rch_Sn132}
Same as the middle panel of Fig.~\ref{fig:Ecorr_Sn132_EHF_Egs_Ecorr_hw12} but for the charge radius.
}
\end{figure}

\begin{figure}[t]
\centering
\includegraphics[width=\columnwidth,clip=]{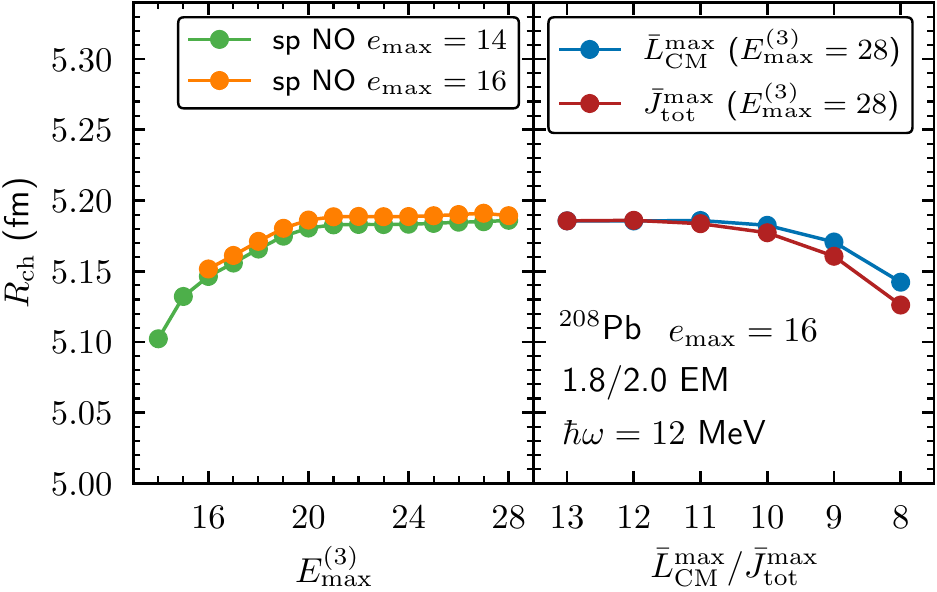}
\caption{
\label{fig:Rch_Pb208}
Same as the middle panel of Fig.~\ref{fig:Ecorr_Pb208_EHF_Egs_Ecorr_hw12} but for the charge radius.
}
\end{figure}

In addition to ground-state energies, we also benchmark results for charge radii. To this end, we evolve the point-proton mean-square radius operator $R_p^2$ in the IMSRG. To obtain the charge radius, we add the relativistic Darwin-Foldy correction $3/(4M^2)=0.033 \, \text{fm}^2$~\cite{Fria97foldyShift} (with $M$ denoting the nucleon mass), the spin-orbit correction $\la r^2 \ra_\text{so}$~\cite{Ong10spinorbit}, as well as the mean-square radii of the proton and neutron $\la r_p^2 \ra = 0.770 \, \text{fm}^2$ and $\la r_n^2\ra=-0.1149 \, \text{fm}^2$, respectively, to obtain the charge radius
\begin{align}
    R_{\text{ch}}^2 =  R_p^2 + \la r_p^2 \ra + \frac{N}{Z} \la r_n^2 \ra + \frac{3}{4M^2} + \la r^2\ra_\text{so} \,.
\end{align}
As for the ground-state energies, we compare results in the Jacobi and single-particle NO frameworks for $^{132}$Sn and $^{208}$Pb in Figs.~\ref{fig:Rch_Sn132} and~\ref{fig:Rch_Pb208}, respectively.

The charge radius operator is known to be already quite well described at the HF level. With an HF charge radius of $R_{\text{ch}} = 4.396 $~fm ($R_{\text{ch}} = 5.141 $~fm) for $^{132}$Sn ($^{208}$Pb), we find only small contributions from the correlation expansion to the charge radius for both NO frameworks. The results in the single-particle NO are converged with respect to $E^{(3)}_\text{max}$ for both cases and show only minor differences when increasing the model space to $e_\text{max}=16$ in Fig.~\ref{fig:Rch_Pb208} for $^{208}$Pb.
We find excellent agreement between both NO approaches and slightly faster convergence in the Jacobi framework with respect to 
$\bar{L}_\text{CM}^\text{max}$ and $\bar{J}_\text{tot}^\text{max}$ compared to the energies in Figs.~\ref{fig:Ecorr_Sn132_EHF_Egs_Ecorr_hw12} and~\ref{fig:Ecorr_Pb208_EHF_Egs_Ecorr_hw12}. 

\section{Summary and Conclusions}
\label{sec:outlook}

In this work, we introduced a new Jacobi NO framework to efficiently and accurately include 3N interactions in \ai{} many-body calculations up to heavy nuclei at the two-body operator level. The Jacobi NO allows to circumvent the need to represent 3N interactions in a single-particle basis in an intermediate step, as required in the traditional NO framework, and hence allows to perform NO in large basis spaces without truncations in energy quantum numbers. The resulting effective interaction in the Jacobi basis explicitly depends on the CM degrees of freedom, characterized by two additional quantum numbers $\bar{L}_\text{CM}$ and $\bar{J}_\text{tot}$ and can be subsequently transformed to a single-particle basis in a straightforward way. We studied the convergence behavior with respect to the new quantum numbers and found excellent agreement for individual matrix elements obtained in the Jacobi and the traditional NO approach for an $^{16}\text{O}$ HF reference state.

We then explored ground-state energies of light, medium-mass, and heavy closed-shell nuclei from $^{16}\text{O}$ to $^{208}\text{Pb}$ using the IMSRG based on the 1.8/2.0 EM interaction of Ref.~\cite{Hebe11fits} and investigated in detail the convergence of the results for both NO frameworks. Excellent agreement was found for the converged energies of $^{16}$O, $^{48}$Ca, and $^{78}$Ni, while for the heavier systems $^{132}$Sn and $^{208}$Pb we found small relative energy differences on the order of about one per mille of the total ground-state energy, which can be traced back to differences in the treatment of the antisymmetrization of the 3N interaction and the employed floating point precision, which we had kept higher in this work. 

In addition, we explored the impact of the $E^{(3)}_\text{max}$ cut used for the HF calculation of the reference state in the Jacobi NO. Even for heavy nuclei like $^{132}$Sn we obtained basically identical results for references states computed using $E^{(3)}_\text{max}=18$ and $E^{(3)}_\text{max}=28$. Thus, at the HF level, smaller $E^{(3)}_\text{max}$ values are needed than for the correlation energy.

Furthermore we observed a systematic increase in the maximum values of $\bar{L}_\text{CM}$ and $\bar{J}_\text{tot}$ required to obtain converged energies as the mass number of the nucleus increases. While $\bar{L}_\text{CM}^{\text{max}}=\bar{J}_\text{tot}^{\text{max}} \approx 5$ is sufficient for $^{16}$O, we need to go to $\bar{L}_\text{CM}^{\text{max}}=\bar{J}_\text{tot}^{\text{max}} \approx 9$ for $^{78}$Ni and eventually to $\bar{L}_\text{CM}^{\text{max}}=\bar{J}_\text{tot}^{\text{max}} \approx 13$ for $^{208}$Pb. This trend is comparable to the single-particle approach, where increased $E^{(3)}_\text{max}$ values are required to obtain converged energies for higher mass numbers. However, going to larger $E^{(3)}_\text{max}$ is significantly more expensive in memory and computing time. Of course, increasing the cuts on the quantum numbers $\bar{L}_\text{CM}^{\text{max}}$ and $\bar{J}_\text{tot}^{\text{max}}$ increases the basis size and hence also the computational complexity of the Jacobi NO framework as well. In its current version the run time for one NO calculation for heavy nuclei like $^{132}$Sn takes approximately one day per $\bar{J}_{\text{tot}}$ channel. Calculations for lighter systems are significantly faster. However, speed-ups may be realized by future optimizations. More refined reference states, such as natural orbital basis states could also be straightforwardly applied in the Jacobi NO. 
Moreover, the framework can be straightforwardly applied to any rotationally invariant density~\cite{Fros21reducedOp}.
Therefore, the Jacobi NO approach can be extended to open-shell nuclei by using, e.g., equal-filling HF or spherical Hartree-Fock-Bogoliubov reference states~\cite{Stro17ENO,Ripoche2020}.

\begin{acknowledgments}
We thank P. Arthuis and T. Miyagi for helpful discussions. We also thank T. Miyagi for providing matrix elements for our benchmark calculations. This work was supported in part by the European Research Council (ERC) under the European Union's Horizon 2020 research and innovation programme (Grant Agreement No.~101020842), the Deutsche  Forschungsgemeinschaft  (DFG,  German Research Foundation) -- Projektnummer 279384907 -- SFB 1245 and Germany's Excellence Strategy -- EXC 2118 PRISMA+ -- 390831469, and by the BMBF Contract No.~05P21RDFNB.
Computations were in part performed with an allocation of computing resources at the Jülich Supercomputing Center.
This work was completed in part at the Helmholtz GPU Hackathon 2021, part of the Open Hackathons program. The authors would like to acknowledge OpenACC-Standard.org for their support.
\end{acknowledgments}

\bibliography{strongint}

\end{document}